\documentclass[preprintnumbers,10pt,a4paper,superscriptaddress,aps,pre,twocolumn]{revtex4-2}

\usepackage{physics}
\usepackage{mathtools}
\usepackage{amsmath}

\usepackage{graphicx}

\usepackage{xcolor}

\usepackage[hidelinks]{hyperref} % this and the next two have to be in this order
\usepackage[capitalise]{cleveref} % automatic referencing

\crefname{plural_equation}{Eqs.}{Eqs.}
\creflabelformat{plural_equation}{(#2\textup{#1}#3)}
\usepackage{xr}

\usepackage{parskip}

\DeclareMathOperator{\E}{E}

\DeclareMathOperator{\covariance}{Cov}
\DeclareMathOperator{\variance}{Var}

\newcommand{\DD}[2][]{\mathinner{\mathcal{D}^{#1}#2}}

\DeclarePairedDelimiter{\avg}{\langle}{\rangle}
\newcommand{\wh}[1]{\widehat{#1}}

\DeclareDocumentCommand\vectorbold{ s m }{\IfBooleanTF{#1}{\boldsymbol{#2}}{\mathbf{#2}}}
\DeclareDocumentCommand\vb{}{\vectorbold}

\newcommand{\nocontentsline}[3]{}
\let\oldaddcontentsline=\addcontentsline
\let\addcontentsline=\nocontentsline

\begin{document}
\title{Interaction networks in persistent Lotka-Volterra communities}
    
\author{Lyle Poley}
\affiliation{Theoretical Physics, Department of Physics and Astronomy, School of Natural Science, The University of Manchester, Manchester M13 9PL, UK}
\author{Tobias Galla}
\affiliation{Instituto de F{\' i}sica Interdisciplinar y Sistemas Complejos IFISC (CSIC-UIB), 07122 Palma de Mallorca, Spain}
\author{Joseph W. Baron}
\affiliation{Laboratoire de Physique de l’Ecole Normale Sup\`{e}rieure, ENS, Universit\'{e} PSL, CNRS, Sorbonne Universit\'{e}, Universit\'{e} de Paris, F-75005 Paris, France}
\date{\today}

\begin{abstract}
      A central concern of community ecology is the interdependence between interaction strengths and the underlying structure of the network upon which species interact. In this work we present a solvable example of such a feedback mechanism in a generalised Lotka-Volterra dynamical system. Beginning with a community of species interacting on a network with arbitrary degree distribution, we provide an analytical framework from which properties of the eventual `surviving community' can be derived. We find that highly-connected species are less likely to survive than their poorly connected counterparts, which skews the eventual degree distribution towards a preponderance of species with low degree, a pattern commonly observed in real ecosystems. Further, the average abundance of the neighbours of a species in the surviving community is lower than the community average (reminiscent of the famed friendship paradox). Finally, we show that correlations emerge between the connectivity of a species and its interactions with its neighbours. More precisely, we find that highly-connected species tend to benefit from their neighbours more than their neighbours benefit from them. These correlations are not present in the initial pool of species and are a result of the dynamics. 
\end{abstract}

\maketitle

\section{Introduction}
The modern discipline of macroecology takes up the ambitious challenge of identifying and understanding the unifying characteristics of ecological communities. Such characteristics include the shapes of abundance distributions, fluctuations in species abundances, and abundance-diversity relationships \cite{grilli2020macroecological, shoemaker2023macroecological, brown1995macroecology}. Of particular interest is the relationship between ecosystem network structure and inter-species relationships \cite{pascualEcologicalNetworksLinking2006,bascompteNestedAssemblyPlant2003, sebastian2015macroecological}. Interaction network structure in real ecological networks has been linked to interspecies competition \cite{bastollaArchitectureMutualisticNetworks2009,bastollaBiodiversityModelEcosystems2005}, stability \cite{Dunne12917,neutelStabilityRealFood2002,lurgiEffectsSpaceDiversity2016,landiComplexityStabilityEcological2018,rohrStructuralStabilityMutualistic2014}, and to the functioning of an ecosystem in the wider biosphere \cite{fuhrmanMicrobialCommunityStructure2009,thompsonFoodWebsReconciling2012}. 

To explain some of these observed relationships, simple models have been suggested (such as the Cascade and Niche models), which have had success in replicating observed patterns in natural foodwebs \cite{Dunne12917,cohen13FoodWebs1989,cohenStochasticTheoryCommunity1990,cohenStochasticTheoryCommunity1997, williamsSimpleRulesYield2000,williamsStabilizationChaoticNonpermanent2004,allesinaGeneralModelFood2008,allesinaPredictingStabilityLarge2015}. The tools of statistical physics and disordered systems are particularly well-suited to aiding in the study of these models, due to their emphasis on deriving universal and emergent phenomena from microscopic interactions. As such, building on the seminal work of Robert May \cite{mayWillLargeComplex1972}, some works have focused on how network structure can influence ecological stability by using random matrix theory \cite{barabasEffectIntraInterspecific2016, allesinaPredictingStabilityLarge2015, grilliModularityStabilityEcological2016, poleyEigenvalueSpectraFinely2023, baron_directed_2022}. However, these models simply posit the structure of the network and interaction coefficients. Therefore, it could be that the hypothesised Jacobian matrix does not correspond to any realistic ecosystem dynamics. 

More recently, dynamic mean-field theory (DMFT) techniques \cite{Galla_2018, buninEcologicalCommunitiesLotkavolterra2017} have been used to examine the statistics of interactions in the surviving communities that result from plausible ecosystem dynamics \cite{barbierFingerprintsHighDimensionalCoexistence2021, barbierGenericAssembltPatterns2018, bunin2016interaction}. It has been shown that intricate correlations between species' interaction coefficients arise in so-called `feasible' model communities, and that these statistics are important for stability \cite{baronBreakdownRandomMatrixUniversality2023b}.

In this work, we seek to understand what kinds of interaction networks are permitted in feasible communities. We present an analytically tractable model in which an initial pool of species interacts according to generalised Lotka-Volterra dynamics. Our interest is in the long-time behavior of the community as it follows these dynamics. The degree distribution of the network on which species initially interact is an input for the model. However, because species can die out during the dynamics, the final network of surviving species is a result of the interactions. In this way, our model captures some salient aspects of the feedback loop between inter-species interactions and the structure of the network on which these interactions take place. The eventual patterns that emerge in the community are a consequence of the fact that the community evolves dynamically; it is feasible by construction. We are thus able to probe the interdependence of interaction network structure and species relationships that characterise feasible communities.

Ultimately, we are able to demonstrate several general trends (for competitive, and stable, communities). First, more highly-connected species are less likely to survive, which skews the degree distribution towards having many species with low connectivity and few species with high connectivity (a pattern observed in nature \cite{williamsSimpleRulesYield2000,dunneFoodwebStructureNetwork2002}). Secondly, species with higher connectivity typically have lower abundance. This in turn means that the average abundance of the neighbour of a randomly selected species is lower than the abundance of a randomly selected species (akin to the so-called `friendship paradox'). Finally, we find that there are correlations between a species' connectivity and its interactions with its neighbours. On average, well-connected species will have more favourable interactions with their neighbours than their neighbours will have with them (and vice versa for poorly-connected species).

The content of this work is organised as follows. In \cref{section:model}, we describe the generalised Lotka-Volterra model and the structure of the interspecies interactions explicitly. We then outline our analytical methods for predicting the behavior of the model in the long-time limit in \cref{section:behaviour_of_the_model}, and we compare our results for the abundance distributions of species in the community to the results of numerical integration. We also analyse stability, and find that network structure can be a stabilising influence in communities with many predator-prey and competitive interactions. In \cref{section:properties_of_the_surviving_community}, we derive a simple expression for the eventual degree distribution of the community, finding that species with low degree become relatively more common, and species with high degree become relatively rare. We also examine the dynamically induced correlations that emerge between species' interactions and the network, and offer biological interpretations for these correlations. We finish by discussing possible extensions to this model and the implications of our results in \cref{section:discussion}.
\section{Model}\label{section:model}
Consider a community of $N$ species interacting according to generalised Lotka-Volterra (gLV) dynamics. These dynamics produce feasible communities by construction (i.e., all abundances remain positive if they are so at $t = 0$). The abundance of species $i$ at time $t$, $x_i(t)$, is determined by the following set of equations 
\begin{align}
      \dot{x}_i(t) = x_i(t) \qty[1 - x_i(t) +  \sum_j A_{ij}\alpha_{ij} x_j(t) ].\label{eq:glv_dynamcis}
\end{align}
The adjacency matrix element $A_{ij}$ encodes the structure of the network on which the species interact. The variable $A_{ij}$ is equal to $1$ if species $i$ and $j$ interact, and is $0$ otherwise. We always impose $A_{ij}=A_{ji}$. The interaction matrix $\alpha_{ij}$ dictates the influence of species $j$ on species $i$, provided that they interact. The values of $\alpha_{ij}$ for which $A_{ij} = 0$ do not play a role in the dynamics. Both the adjacency matrix $\vb A$ and interaction matrix $\vb *\alpha$ are random matrices. They are selected independently of each other, and are fixed throughout the dynamics.

We construct the matrix $\vb A$ according to the random configuration model \cite{bollobasProbabilisticProofAsymptotic1980,newmanNetworks2018}  (also known as the Chung-Lu model \cite{chungConnectedComponentsRandom2002}). This generalises the often-used Erd\"os Reyni network to incorporate an arbitrary degree distribution, which we write as $p_k$. To generate an instance of the network, we first draw the degree of each node independently from $p_k$. With this degree sequence $\{{k_i}\}$, we set each pair $(A_{ij}, A_{ji})$ to one with probability $k_ik_j/(dN)$, and set the pair to zero otherwise. For sufficiently large $N$, this construction will produce networks with the desired degree distribution. To ensure that the probability of connection is well defined, we require $k_ik_j < dN$ for all $(i, j)$. 

For simplicity, we will assume that the degree distribution is a uniform distribution, with width $w = k_\text{max} - k_\text{min}$ and average degree $d$, although our approach applies to any degree distribution. In our examples, both $d$ and $w$ are proportional to $N$. In the case of the uniform degree distribution, the condition $k_ik_j < dN$ for all $(i,j)$ is equivalent to $(w/2 + d)^2 < dN$.

To construct an instance of the interaction matrix $\vb *\alpha$, pairs of elements $(\alpha_{ij}, \alpha_{ji})$ are drawn identically and independently from a probability distribution with the following statistics 
\begin{align}
      \avg*{\alpha_{ij}}_\alpha &= \frac{\mu}{d}, \nonumber\\
      \variance\qty(\alpha_{ij})_\alpha &= \frac{\sigma^2}{d}, \nonumber\\
      \covariance\qty(\alpha_{ij}, \alpha_{ji})_\alpha &= \frac{\gamma\sigma^2}{d}.\label[plural_equation]{eq:interaction_statistics}
\end{align}
The statistics of the interaction matrix are scaled with the factor of $1/d$ to ensure that the ensemble-averaged interaction strength between species in the community is $\sum_{ij}\avg{A_{ij}\alpha_{ij}}_{A, \alpha}/N = \mu$, and similarly for the variance $\sum_{ij}\avg{(A_{ij}\alpha_{ij} - \mu/N)^2}_{A, \alpha}/N = \sigma^2$, which is commonly the case in fully connected versions of the model \cite{Galla_2018,buninEcologicalCommunitiesLotkavolterra2017}. Our definition of the model parameters requires $\sigma^2 > 0$ and $\abs{\gamma} \leq 1$.

The correlation coefficient $\gamma$ controls the symmetry of interactions $\alpha_{ij}$ and $\alpha_{ji}$ in the original community, with $\gamma = 1$ for symmetric interactions ($\alpha_{ij}=\alpha_{ji}$ for all $i\neq j$), and $\gamma = -1$ for antisymmetric deviations from the mean [$\alpha_{ij} -\mu/d=-(\alpha_{ji}-\mu/d)$ for all $i\neq j$]. If we were to further assume that the pairs $(\alpha_{ij}, \alpha_{ji})$ were jointly Gaussian distributed, then $\gamma$ has a simple relationship to the proportion  $p$ of interactions in the community that are of predator-prey type: $\gamma = \cos(\pi p)$ \cite{poleyGeneralizedLotkaVolterraModel2023}. Generically, $\gamma$ is a decreasing function of the proportion of predator-prey-type links in the community.
\section{Behavior of the model}\label{section:behaviour_of_the_model}
\begin{figure}
      \includegraphics[width=0.48\textwidth]{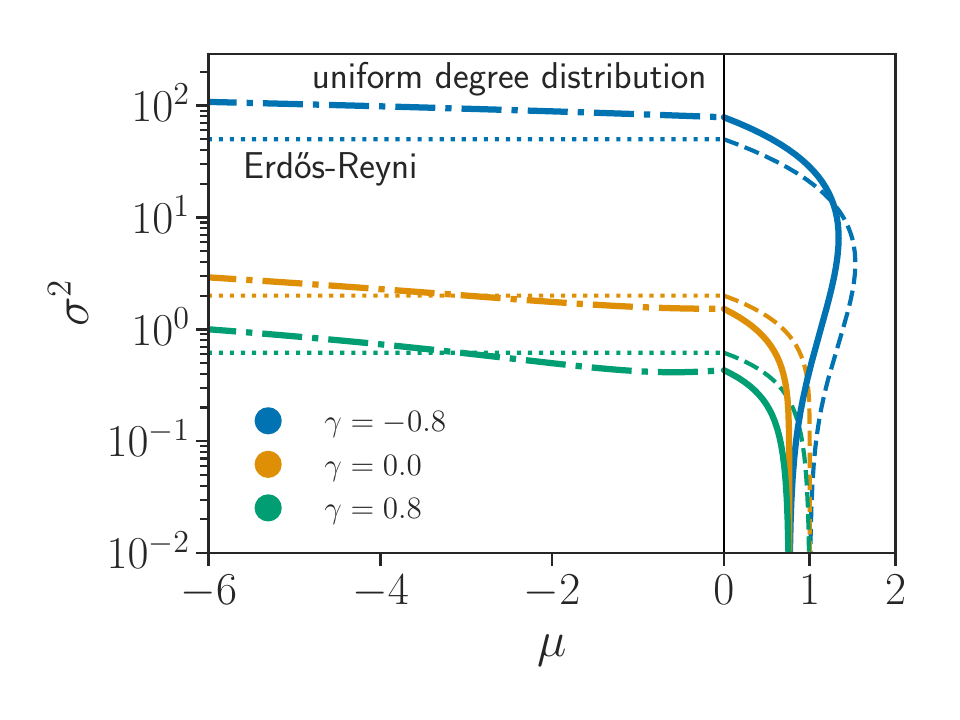}
      \caption{Stability plot for different values of $\gamma$ for the network model as described in \cref{section:model} (solid and dot-dashed lines), and for the same model on an Erd\H{o}s-Reyni graph (dashed and dotted lines) with the same average degree per species. Solid (dashed) lines indicate the transition between the stable and diverging abundance phases, and dashed (dotted) lines indicate the onset of linear instability. Areas to the left of the solid curves, and below dashed curves are stable. For all curves, the average degree is $d = N/4$. For solid and dashed lines the width is $w = 0.5N$.}
      \label{fig:stability}
\end{figure}
\subsection{Dynamical mean-field theory and phase diagram}
Depending on the model parameters ($\mu, \sigma, \gamma$ and the degree distribution $p_k$), the dynamics in \cref{eq:glv_dynamcis} exhibit three distinct phases

As in existing gLV models with random all-to-all interactions \cite{Galla_2018, buninEcologicalCommunitiesLotkavolterra2017, altieri2021properties}, there is a phase in which, for a fixed interaction matrix, the dynamics converge to a unique equilibrium independently of the initial abundances. Secondly, there is a phase in which there are multiple stable fixed points for any given interaction matrix, or the system can remain volatile indefinitely. Finally, species abundances diverge in a third phase. 

The three phases are separated from one another by the onset of a linear instability and, secondly, by the onset of diverging abundances. 
We give an overview of our results for the phase diagram of the model in \cref{fig:stability}. As discussed in Refs.~\cite{mallminChaoticTurnoverRare2024, garcialorenzana2022well}, there are other possible regimes when the interactions do not scale with the system size as above. However, we do not consider these regimes in this work.

The focus of our analytical work is on the properties of the phase in which the dynamics always converge to a unique equilibrium (independent of the initial species abundances). To this end, we employ a generating-functional method, which has its roots in the physics of disordered systems \cite{mezard1987,De_Dominics1978,janssenLagrangeanClassicalField1976b,martinStatisticalDynamicsClassical1973a}, to derive dynamical mean-field equations for the time evolution of the community. Dynamical mean-field theory has been successfully applied to ecological models since the work of Ref.~\cite{1overf_noise} in the context of replicator equations and since the work of Ref.~\cite{Galla_2018} in the context of the gLV equations (see also \cite{roy2019numerical}). 

In most previous models, the DMFT formalism produces a single effective process, which describes the dynamics of a `typical' species abundance. The statistics of the evolution of this effective species mirror those of the entire community. In this work, because species in the original community are distinguishable by their degree, there is an effective process representing the typical behavior of species of each possible degree. Our treatment here follows that used to analyse a previous model in which species were distinguished not by their degree, but by their position in a hierarchy \cite{poleyGeneralizedLotkaVolterraModel2023}.
\begin{figure}
      \includegraphics[width=0.48\textwidth]{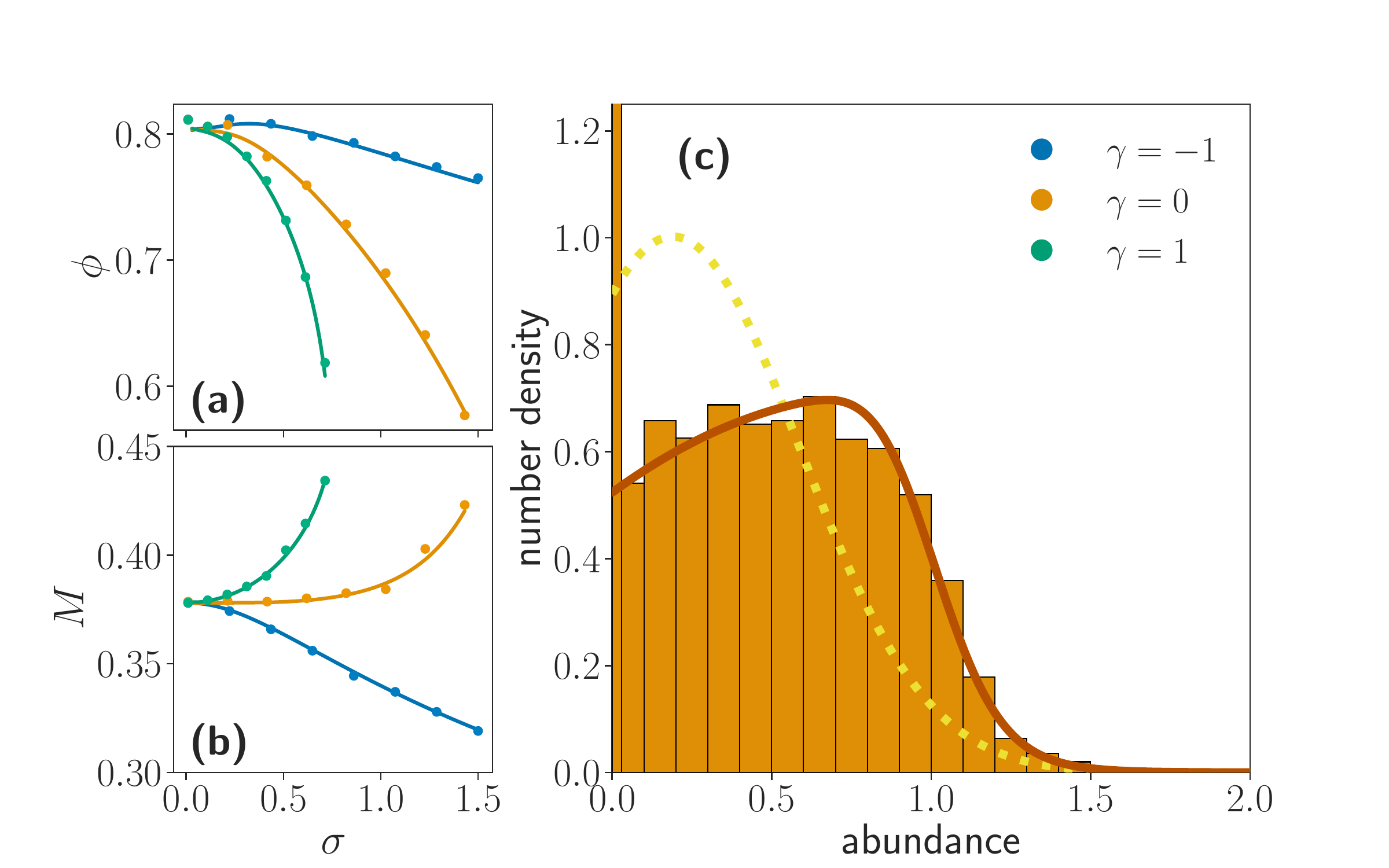}
      \caption{(a) and (b): Overall survival probability $\phi$ and mean abundance $M$ in the community for varying variance of interaction strength $\sigma$. (c): The abundance distribution of species with degree $k$ is a clipped Gaussian, and the average of all such curves gives the abundance distribution of the community as a whole (solid orange curve, see \cref{appendix:abundance_distribution} for details). The corresponding prediction from the Erd\H{o}s-Reyni model with the same mean, variance, correlation, and average degree $d$ is shown as a yellow dotted curve. For all plots, the degree distribution is uniform, and the parameters are $\mu = -3$, $w = 0.45N$ and $d = N/4$ and $\gamma$ is as indicated in the legend. In (a) and (b), $N = 1000$ and markers are the average over $10$ runs of the dynamics. In (c), $\sigma = 1$ and $N = 5000$, and the histogram is the result of single run of the dynamics. The large spike at zero abundance is due to species which have died.}
      \label{fig:abundance_distribution}
\end{figure}
\subsection{Characterisation of the unique-equilibrium phase}\label{section:characterisation_of_UEP}
In the unique-equilibrium phase, we can use the DMFT equations to find the abundance distribution of species that have a particular degree $k$ in the original community. The derivation of the DMFT equations, as well as the analysis of the fixed point, can be found in \cref{appendix:dmft}. 

The abundance $x_k^\star$ of a typical species with degree $k$ at the fixed point is a random variable with a clipped Gaussian distribution. The fixed-point abundances satisfy $x_k^\star(z) = \max[0, x_k^{\star+}(z)]$, where the non-zero abundances satisfy
\begin{align}
      x^{\star+}_k(z) =
      \frac{1 + \mu \frac{k}{d} \sum_{k'} p_{k'}\frac{k'}{d}M_{k'} + z\sigma \sqrt{\frac{k}{d}\sum_{k'} p_{k'}\frac{k'}{d}q_{k'}} + h}{1 - \gamma\sigma^2\frac{k}{d}\sum_{k'} p_{k'}\frac{k'}{d}\chi_{k'}}. \label{eq:stationary_abundance}
\end{align}
The quantity $z$ is a zero-mean, unit-variance Gaussian random variable, and $h$ is an external field used to define the response function $\chi_k$ below. At the end of the calculation, and in all simulations, we set $h=0$.  The quantities $M_k$ and $q_k$ are the first and second moments of the distribution of $x^\star_k$ respectively. These objects are determined self-consistently from their definitions
\begin{align}
      M_k &= \int_{x_k^\star>0} \dd{z} f(z)~ x^\star_k(z), \nonumber \\
      q_k &= \int_{x_k^\star>0} \dd{z} f(z)~ x^\star_k(z)^2, \nonumber \\
      \chi_k &= \int_{x_k^\star>0} \dd{z} f(z)~ \pdv{x^\star_k(z)}{h}, \label[plural_equation]{eq:fixed_point_equations}
\end{align}
where $f(z) = \exp(-z^2/2) /\sqrt{2\pi}$ is the probability density function of the standard normal distribution. For given model parameters, we can solve \cref{eq:stationary_abundance,eq:fixed_point_equations} numerically to yield the values of the $M_k, q_k$ and $\chi_k$, for $k=k_\text{min},\dots,k_{\rm max}$ (where we write $k_{\rm min}$, $k_{\rm max}$ for the lowest and highest degree in the network respectively). These quantities in turn yield the abundance distributions, i.e., the distributions for the different $x_k^\star$. We also define the probability of survival for species that have degree $k$ in the original community
\begin{align}
      \phi_k = \int_{x_k^\star>0}\dd{z} f(z),
\end{align}
as well as the community wide abundance and survival probability 
\begin{align}
    M = \sum_kp_kM_k,\qquad \phi = \sum_kp_k\phi_k.
\end{align}

Figure \ref{fig:abundance_distribution} confirms the validity of the fixed point solution from Eqs.~(\ref{eq:stationary_abundance}) and (\ref{eq:fixed_point_equations}). The prediction for the average abundance, survival rate and total abundance distribution across the community match the results of simulations. We also show the prediction for the abundance distribution from a theory which does not take the full network structure into account, but instead assumes an Erd\H{o}s-R\'eyni (ER) network with the same average degree $d$. The Erd\H{o}s-R\'eyni network is obtained by setting $p_k = \delta_{k,d}$, where $\delta_{k, d}$ is the Kronecker delta. Although there is degree heterogeneity in this network, it is of the order $1/d$ \cite{baron_directed_2022}. As we can see from \cref{fig:degree_distribution}, the ER graph does not share the same abundance distribution as when $p_k$ is uniform distribution, confirming the importance of degree heterogeneity in the theory.

\subsection{Onset of instability}
The analytical results presented in \cref{section:characterisation_of_UEP} are only valid in the phase with a unique equilibrium. In \cref{appendix:stability}, we find the boundary of this stable regime in terms of the parameters of the model. 

The onset of the diverging phase for given model parameters ($\mu, \sigma, \gamma$ and $p_k$) is found by solving the fixed point equations (\ref{eq:fixed_point_equations}), together with the additional condition that the community average abundance $M$ diverges.

To identify the point at which the dynamics become linearly unstable, we consider a small random perturbation to the fixed point abundances $x^\star_k(t) = x^\star_k + \epsilon y_k(t)$. In the stable regime, this perturbation will decay to zero. In the unstable phase, the abundances will not in general return to $x^\star_k$ after being perturbed. In \cref{appendix:stability}, we find that such a perturbation will eventually decay to zero provided the following condition holds
\begin{align}
      \frac{\sigma^2}{d^2}\sum_kp_k\frac{k^2\chi_k^2}{\phi_k} < 1. \label{eq:linear_stability}
\end{align}
That is to say, the system is stable against linear perturbations if this inequality is satisfied, and is not otherwise.

Solving the fixed point equations [\cref{eq:fixed_point_equations}] simultaneously with the condition obtained from setting the left-hand side of the inequality in \cref{eq:linear_stability} equal to one gives us the boundary of the stable and linearly unstable phase. In the fully connected system, the condition for the onset of the linear instability reduces to $\sigma^2\phi(1 + \gamma)^2 < 1$, which has been derived previously using both DMFT \cite{Galla_2018}, and the static cavity method \cite{buninEcologicalCommunitiesLotkavolterra2017}.
\begin{figure}
      \includegraphics[width=0.48\textwidth]{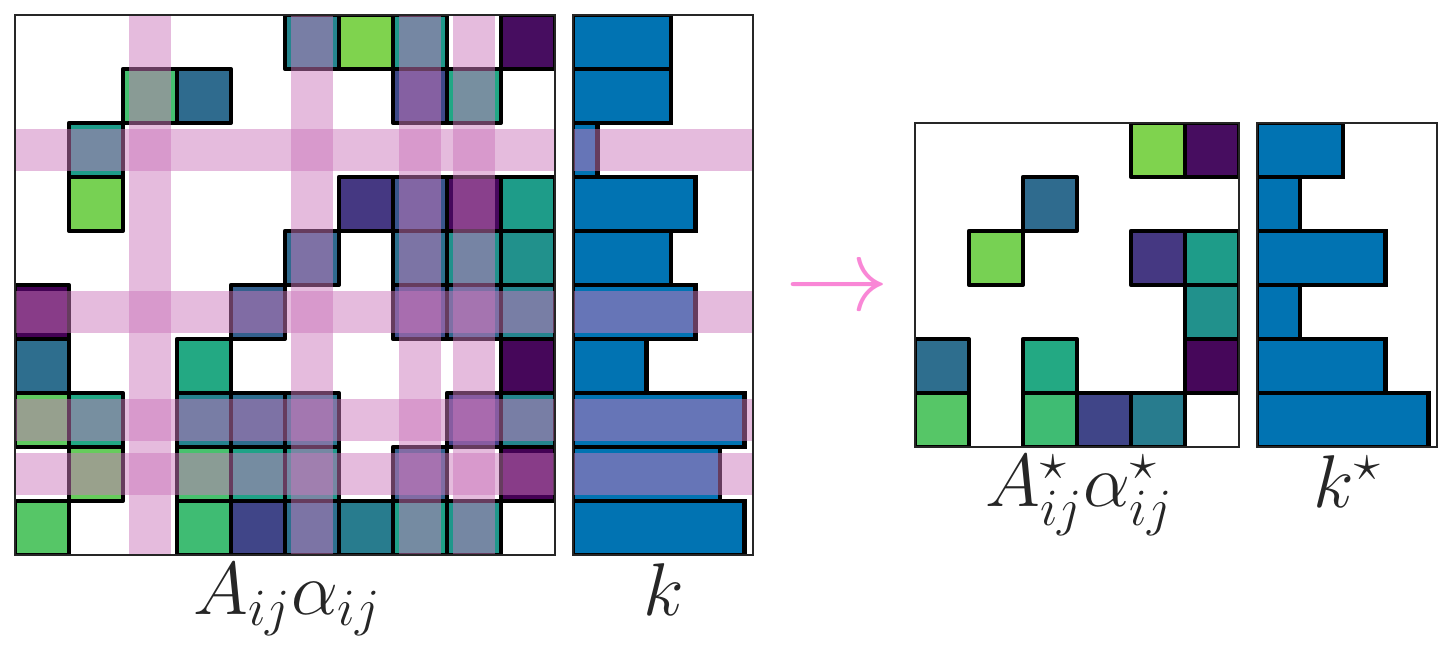}
      \caption{Structure of the interaction matrix and corresponding degree sequence in the original pool of species vs their counterparts in the surviving community. Pink strips indicate which species go extinct in the course of the dynamics, and are hence removed from the community. In this example (counting from the top row of $A_{ij}\alpha_{ij}$), $S^\star = \{1, 2, 4, 5, 7, 10\}$. The degree sequence among the survivors is not simply the sequence of original degrees restricted to surviving species, because some of the survivors' interaction partners also die out. Generally, the degree of any extant species in the surviving community will be lower than its original degree.}
      \label{fig:reduced_network_illustration}
\end{figure}
\section{Properties of the surviving community}\label{section:properties_of_the_surviving_community}
We have established the analytical theory for describing the overall properties of the surviving community, as well as the conditions under which this theory is valid. We now turn our attention to underlying statistics of the network and the interactions of surviving community. Specifically, in this section we will quantify how the survival rates, abundances and interaction strengths between species depend on their connectivity. 

Throughout this section, we write $\vb A^\star, k^\star_i, p^\star, \vb*\alpha^\star$ for the adjacency matrix, the degree (connectivity) of a species $i$, the degree distribution, and the matrix of interaction strengths in the surviving community respectively. We also write $N^\star$ for the number of species in the surviving community, and $S^\star$ for the set of all persisting species. We emphasise that $\vb A^\star, k_i^\star, p^\star, \vb*\alpha^\star$ are not the same as the corresponding quantities in the original community $\vb A, k_i, p, \vb*\alpha$. The differences between the two are the result of the interaction-dependent species extinctions that occur during the course of the dynamics. The relationship between $A_{ij}\alpha_{ij}$, $k_i$ and $A^\star_{ij}\alpha^\star_{ij}$, $k^\star_i$ is illustrated in \cref{fig:reduced_network_illustration}.

\subsection{Structure of the network}
\label{section:surviving_network}
In this section we will use the statistics of $A^\star_{ij}$ to find expressions for the degree of a species in the \textit{surviving} community, given its degree in the original community, as well as the degree distribution in the surviving community. One crucial observation that will aid us in doing this is the following. The probability of any two species interacting in the surviving community (i.e. conditioned on the survival of both species) is $kk'/(dN)$, where $k, k'$ are the species degrees in the \textit{original} community. This is because, to leading order in $1/d$, the survival of different species can be treated as independent events. We discuss this in more detail in \cref{appendix:surviving_network}. 

One notes that although the conditional probability that species interact \textit{given} their survival is trivially related to their interaction probability in the original pool (they are equal), the probability that both species actually survive is dependent on their respective degrees. This leads to non-trivial changes in the network structure.

\begin{figure}
      \includegraphics[width=0.48\textwidth]{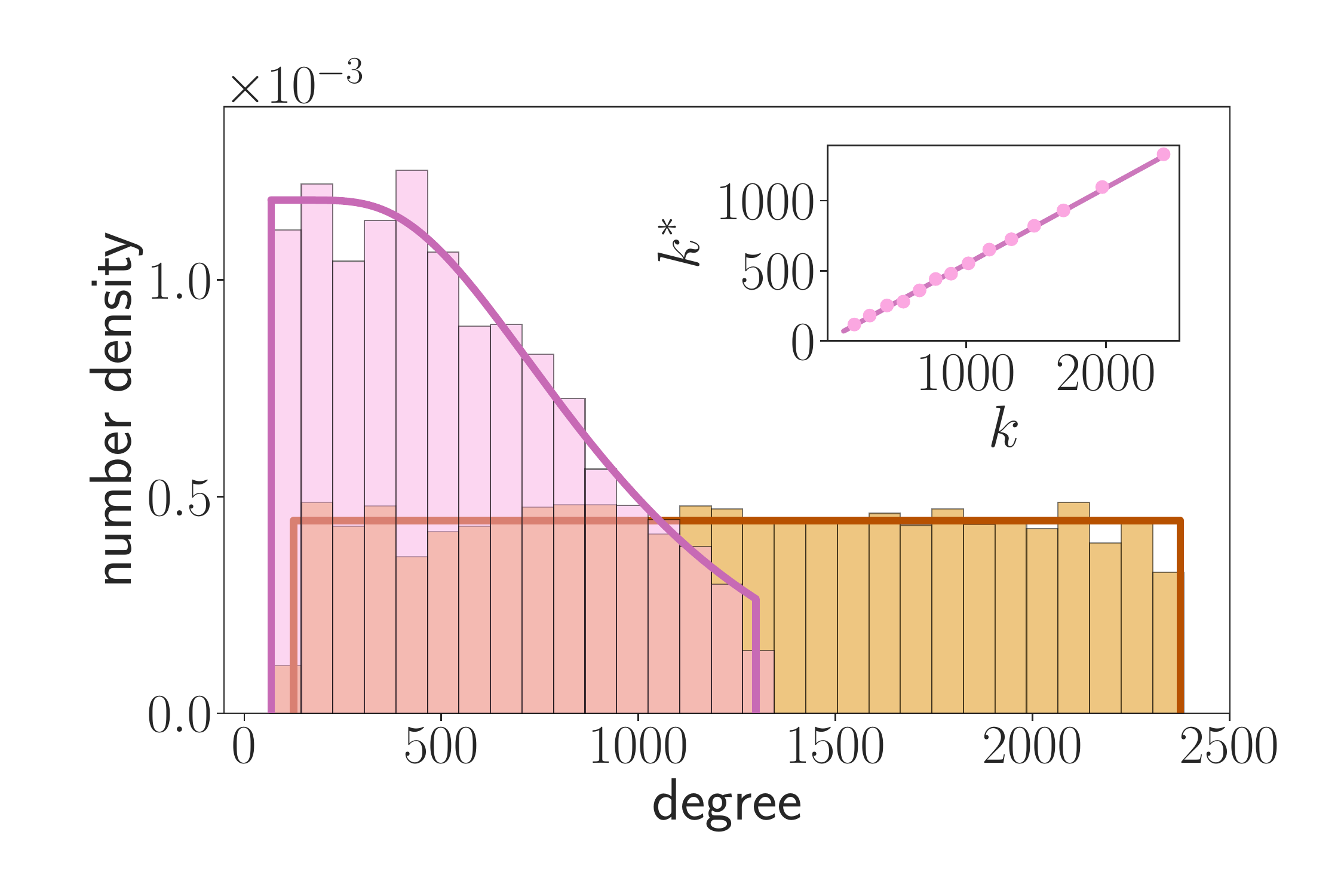}
      \caption{Degree distribution in the surviving community (tall pink distribution) and the original species pool (flat orange distribution). The degree distribution in the original community is uniform, while the degree distribution in the surviving community is given by \cref{eq:surviving_degree_distribution}. Parameters are the same as \cref{fig:abundance_distribution}(c). The inset shows the degree sequence in the surviving community as a function of the degree sequence in the original community. Bars and markers are computed from a single run of the dynamics with $N = 5000$. To avoid too many points in the inset, we only display markers for one in every 250 species in the surviving community. No average is taken.}
      \label{fig:degree_distribution}
\end{figure}

The expected degree of a surviving species, given its degree in the original community, can be computed from our expression for the probability of any two species interacting in the surviving community (see \cref{appendix:degree_distribution} for details). We find, for the expected degree,  
\begin{align}
    \E[k^\star_i\mid k_i, i\in S^\star] = rk_i,\label{eq:surviving_degree_sequence}
\end{align}
where $\E[\dots]$ denotes the combined average over $\vb A$ and $\vb* \alpha$, and where $r = \sum_kp_k\phi_k k/\sum_kp_k k$ is the survival probability of the neighbours of an arbitrarily chosen species in the original community. In \cref{appendix:surviving_network}, we further show that the variance $\E[(k^\star_i)^2\mid k_i, i\in S^\star] - \E[k^\star_i\mid k_i, i\in S^\star]^2$ is sub-leading in $1/d$. Hence, the probability distribution of species' degrees in the surviving community, given that they had degree $k$ in the original community, is highly concentrated around the mean value given in \cref{eq:surviving_degree_sequence}. This can be seen in \cref{fig:degree_distribution} inset, which shows a scatter plot of the points $(k_i^\star, k_i)$ for $i \in S^\star$ for one particular instance of $A$ and $\alpha$ (i.e. there is no average performed). We see an almost perfect linear relationship between $k^\star$ and $k$ with very little fluctuation. From \cref{eq:surviving_degree_sequence}, the gradient of this line is $r$.

Using \cref{eq:surviving_degree_sequence}, we can thus find a compact expression for the degree distribution in the surviving community $p^\star_{k^\star}$. We also use the fact that, for many possible degrees and large $N$, the integer spacing between different possible degrees effectively becomes a continuum. For this reason, we write $P^\star(k^\star/M) = Mp^\star_{k^\star}$ and $P(k/M) = Mp_{k}$, where $M$ is the number of different degrees in the community (the number of distinct values of $k_i$). We can express $P^\star(\kappa^\star)$ (where $\kappa^\star = k^\star/N$ is a variable between $0$ and $1$) in terms of the original degree distribution $P(\kappa)$ and survival rate $\phi(\kappa)$ as
\begin{align}
    P^\star(\kappa^\star) = \frac{1}{\phi r}\phi\qty(\frac{\kappa^\star}{r})P\qty(\frac{\kappa^\star}{r}), \label{eq:surviving_degree_distribution}
\end{align}
where $\phi(k/M) = \phi_k$. This can be understood as follows: the probability $P^\star(\kappa^\star)$ that a randomly selected species in the surviving community has degree $k^\star$ is proportional to the product of the  probabilities that a randomly selected species in the initial pool has degree $k=k^*/r$ and that this species survives [$P(\kappa) \phi(\kappa)$]. The factor of $1/(\phi r)$ is a normalisation constant, ensuring that $\int P^\star(\kappa^\star)\dd{\kappa^\star} = 1$ (see \cref{appendix:degree_distribution} for details).

In \cref{fig:degree_distribution}, we show the effect of the dynamical selection on a community interacting on a network that initially has a uniform degree distribution. It is clear that, relative to the initial degree distribution, there are more species with low degree than with high degree in the surviving community. This is driven by the fact that highly connected species in the original community are less likely to survive than species with low degree in competitive communities (i.e. for $\mu < 0$, see \cref{appendix:phik_Mk_decreasing_functions} and the next section for details).

\subsection{Survival rates and abundances as functions of degree}\label{section:Mk_and_phik} 

\begin{figure}
    \includegraphics[width=0.48\textwidth]{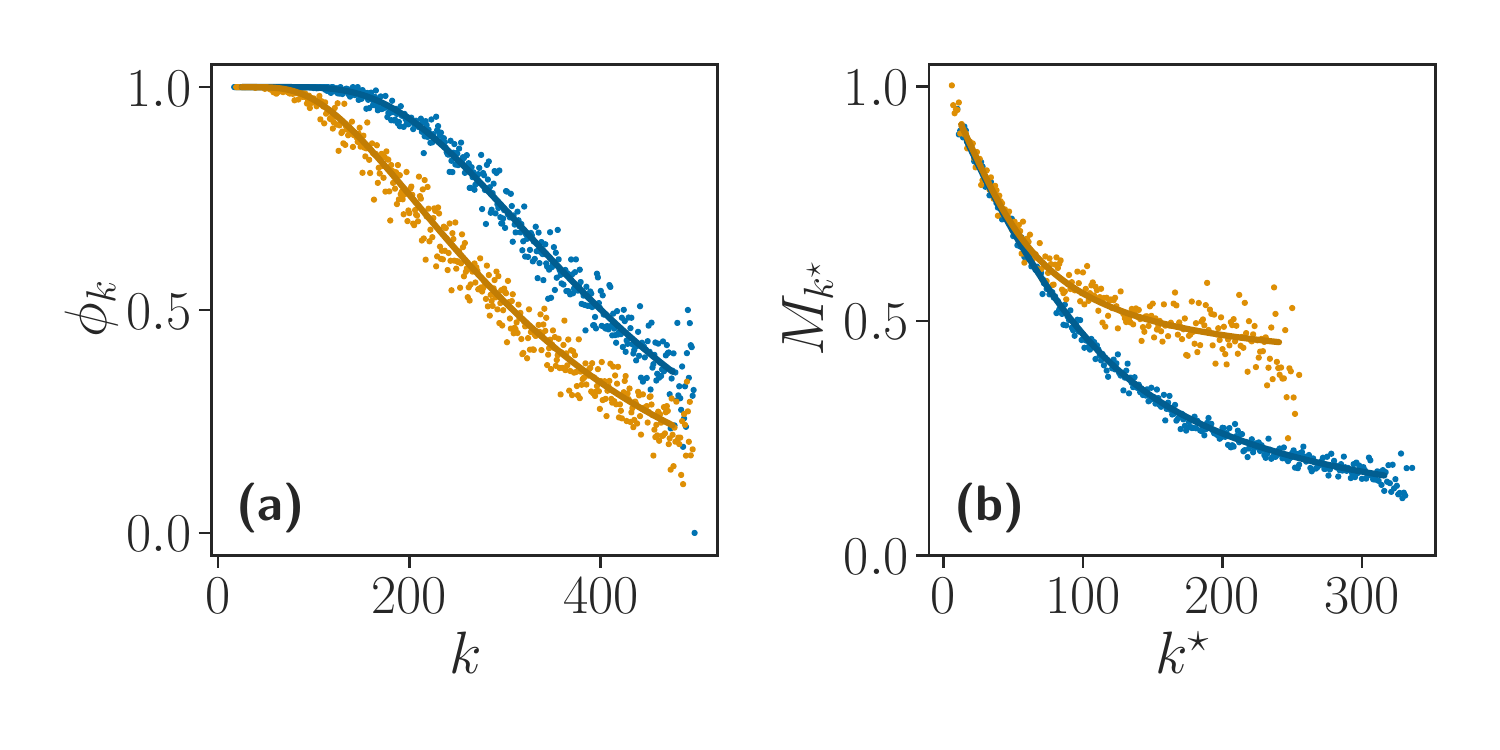}
    \caption{(a) Survival rate as a function of degree in the original community. (b) Abundance as a function of degree in the surviving community. The survival rate is written as a function of degree in the original community, rather than as a function of degree in the surviving community. In contrast, the abundance of a species with given degree is an observable that requires only information about the surviving community, hence we plot it as a function of $k^\star$ rather than $k$. The parameters are $\mu = -3, \sigma=1.2, d=N/4, w = 0.45N, N = 2000$, and the values of $\gamma$ are the same as in \cref{fig:abundance_distribution}. In Panel (b), the curves terminate at different values of $k^\star$ because of the differences in the limit of the degree sequence $k^\star$ in the surviving community. Markers are the average over $200$ runs of the dynamics, and the solid lines are analytical predictions.}
    \label{fig:phik_and_Mk_correlations}
\end{figure}

As we see in \cref{fig:phik_and_Mk_correlations}(a), species with higher initial degree are less likely to survive.  This can be understood in broad terms from \cref{eq:stationary_abundance}. The abundance $x_k^\star$ is a random variable drawn from a clipped Gaussian distribution. That is, if the Gaussian variable $z$ is such that the RHS of \cref{eq:stationary_abundance} is negative, the species does not survive. Given that the factor multiplying $z$ is proportional to $\sqrt{k}$ and that the factor multiplying $\mu<0$ is proportional to $k$, we see that as $k$ increases, it is more likely that the RHS of \cref{eq:stationary_abundance} is negative. Hence, a higher fraction of species go extinct for higher $k$. This is always the case for $\mu < 0$ (see \cref{appendix:phik_Mk_decreasing_functions}).

\cref{fig:phik_and_Mk_correlations}(b) shows the expected abundance $M_{k^\star}$ of species as a function of their degree in the surviving community, which in the case shown is also seen to be a decreasing function of $k^\star$. This is not always guaranteed to be the case however, even for $\mu<0$. With that being said, in \cref{appendix:phik_Mk_decreasing_functions}, we show that the region in parameter space for which the system is stable, where $\mu < 0$, and for which $M_{k^\star}$ and $k^\star$ are positively correlated is small. Hence, for $\mu < 0$, only a small range of parameters could give rise to a community in which $M_{k^\star}$ is an increasing function of $k^\star$. This general trend can once again be understood from \cref{eq:stationary_abundance}, where we see that the term proportional to $\mu$, which determines the typical value of $x^\star_{k}$, is also proportional to $k$.

The fact that $\phi_k$ and $M_{k^\star}$ are decreasing functions of degree has consequences for the relationship between species and their neighbours in the community. In \cref{section:surviving_network}, we introduced the probability of survival of the neighbours of a species in the original community $r = \sum_kp_k\phi_kk/\sum_kp_kk$. We now show that if $\phi_k$ decreases with $k$, then $r < \phi$. That is, the probability of survival of the neighbours of a species is lower than the overall probability of species survival. 

To see this, we first observe that if $\phi_k$ decreases with $k$, then the covariance of $\phi_k$ and $k$ (computed with respect to the degree distribution $p_k$) must be negative. That is, $\sum_k p_k \phi_kk - (\sum_{k}p_{k}k)(\sum_{k'}p_{k'} \phi_{k'}) < 0$. We arrive at our claim after dividing both sides of the inequality by $\sum_kp_kk$ and recalling that $\phi = \sum_kp_k\phi_k$. 

By an identical argument, we can conclude that if $M_{k^\star}$ is a decreasing function of $k^\star$, then the average abundance of the neighbours of a species in the community is lower than the average abundance in the surviving community as a whole. Following the colloquial statement of the famous friendship paradox, `your friends are more popular than you are', we could say that ``species' neighbours are less populous than they are''.
\begin{figure*}
      \includegraphics[width=0.96\textwidth]{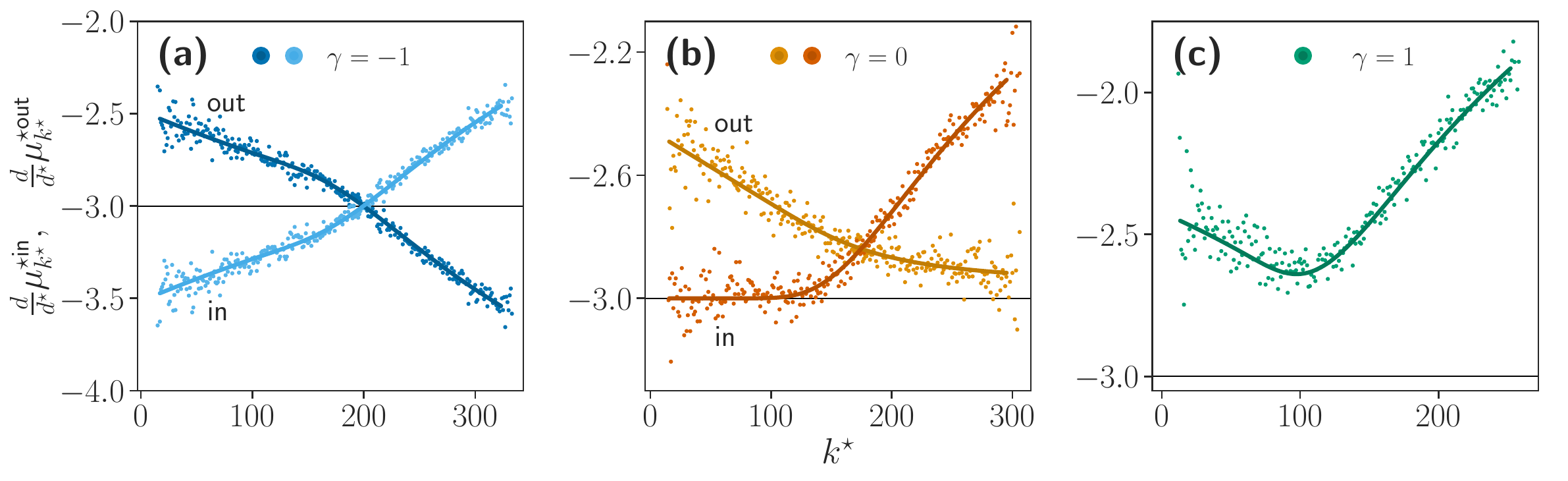}
      \caption{Average `in' and `out' interaction strength as a function of degree in the surviving community with (a, b, c) corresponding to $\gamma = (-1, 0, 1)$ respectively. In (a) and (b), the curves with positive gradient show the`in' interaction strengths and the curves with negative gradient show the `out' interaction strengths. In (c), the `in' and `out' interactions are exactly equal. A horizontal black line has been added to all panels to indicate the mean interaction strength in the original community. Parameters are $\mu=-3, \sigma=0.6, d=N/4, w=0.45N$, $N=2000$. Markers are the average of $200$ runs of the simulation. Solid lines are analytical predictions (see \cref{appendix:interaction_statistics} for the explicit expressions).}
      \label{fig:interaction_degree_correlations}
\end{figure*}
\subsection{Interaction strengths in the surviving community}
It is known that intricate correlations between interaction coefficients, which are not present in the original community, emerge in the surviving communities of fully-connected gLV systems \cite{bunin2016interaction,buninEcologicalCommunitiesLotkavolterra2017,baronBreakdownRandomMatrixUniversality2023b}. In this section, we show that in network gLV systems, the dynamics also induce correlations between the degree of a species and its interaction coefficients in the surviving community, even though there are no such correlations in the initial community.

To quantify this effect, we characterise the strength of interactions `coming into' and `going out of' a species with degree $k$ in a general network as
\begin{align}
      \mu^{\text{in}}_k &= \frac{d}{N_k}\sum_{i \in S_k}\frac{1}{k}\sum_{j}A_{ij}\alpha_{ij}, \nonumber\\
      \mu^{\text{out}}_k &= \frac{d}{N_k}\sum_{i \in S_k}\frac{1}{k}\sum_{j} A_{ji}\alpha_{ji}, \label{eq:muinmuout}
\end{align}
where we write $S_k$ for the set of species that have degree $k$ and $N_k$ for the number of species in this set. Because the network and interaction strengths are independent in the original community, the ensemble average in and out interaction strengths are both equal to $\avg{\mu_k^\mathrm{in/out}}_{A, \alpha} = \mu$ for any value of $k$. However, as \cref{fig:interaction_degree_correlations} demonstrates, when we measure these same quantities in the surviving community [i.e. $\mu^{\star\text{in}}_{k^\star} = (d^\star/N^\star_{k^\star}) \sum_{i\in S^\star_{k^\star}} \sum_{j\in S^\star} A^\star_{ij}\alpha^\star_{ij}/k^\star$ and similar for $\mu^{\star\text{out}}_{k^\star}$], we find that they are functions of the degree $k^\star$. 

We can understand the relationship between the in/out interaction strengths and connectivity by examining \cref{fig:interaction_degree_correlations}(b). In this case, $\gamma = 0$, and therefore there is no imposed correlation between the `in' ($\alpha_{ij}$) and `out' ($\alpha_{ji}$) interaction coefficients, so we are better able to disentangle the effects at play. 

Let us begin with the incoming interactions. We see that the average incoming interaction strength increases with degree $k^\star$. We attribute this primarily to the differing survival rate of species, depending on degree. We can see this as follows. Almost all species with small $k$ survive (i.e. $\phi_k \approx 1$), and so we expect the average incoming interactions to be same in the surviving and initial communities for species with small $k^\star$ (i.e. $\mu_{k^\star}^{\star\mathrm{in}} \approx \mu$). As we increase the value of $k^\star$, the survival rate of the species decreases (see \cref{fig:phik_and_Mk_correlations}). Because species with the most favourable interactions survive, a `selection bias' is introduced for species with higher degree, hence the upwards trend of $\mu_{k^\star}^{\star\mathrm{in}}$ with $k^\star$.

The average outgoing interaction strength exhibits the opposite trend to the incoming interaction strength in \cref{fig:interaction_degree_correlations}(b). Low-degree species have more positive outgoing interactions than high-degree species. This is because low-degree species have higher abundance (see \cref{section:Mk_and_phik}), and therefore have a greater impact on the probability of their neighbours' survival. Hence, the neighbours of low-degree species are under relatively high selection pressure compared to neighbours of species with high degree (which have comparatively low abundance). As was the case with the incoming interactions, it is those species who interact more favourably that survive. Because there is less selection bias for the neighbours of species with higher $k^\star$, we see a corresponding decrease in $\mu^{\star\text{out}}_{k^\star}$ with $k^\star$. 

The trends in the other panels of \cref{fig:interaction_degree_correlations} can be understood as a kind of superposition of the trends in panel b, since a non-zero $\gamma$ connotes a correlation between the in- and out-interactions of a species. For example, when $\gamma = 1$, the outgoing and incoming interactions must be identical. Hence, $\mu_{k^\star}^{\star\mathrm{out}} = \mu_{k^\star}^{\star\mathrm{in}}$ and the corresponding curve in \cref{fig:interaction_degree_correlations}(c) is seen to be an `average' of the upwards and the downwards trends in panel (b). On the other hand, for $\gamma = -1$, the fluctuations in the in- and out-interactions are exactly the negative of each other [i.e. $\alpha_{ij} - \mu/d = -(\alpha_{ji} - \mu/d)$], hence the mirror-image effect in \cref{fig:interaction_degree_correlations}(a).

Although we have provided a qualitative rational of the trends in Fig. \ref{fig:interaction_degree_correlations}, we are also able to provide direct quantitative evidence of their causation using the cavity method (in a similar way to Refs. \cite{bunin2016interaction, baronBreakdownRandomMatrixUniversality2023b}). That is, we see directly how the survival bias of species affects the incoming interactions, and how the species abundance affects the outgoing interactions. The cavity method also yields the analytical results in \cref{fig:interaction_degree_correlations}. The details are technical, and so we direct the interested reader to \cref{appendix:interaction_statistics} for more information.

\section{Discussion}\label{section:discussion}

In this work, we have studied an extension of the popular generalised Lotka-Volterra equations by incorporating random network structure with an arbitrary degree distribution. We have found that the network and interaction statistics of the surviving community differ greatly from those of the original community. This demonstrates that the condition of feasibility, which linear models cannot guarantee, is a strong constraint on the structure of ecological networks. To derive our central results, we extended the usual dynamical mean-field theory for generalised Lotka-Volterra dynamics to describe heterogeneous interaction statistics, in a similar way to our previous work in Ref. \cite{poleyGeneralizedLotkaVolterraModel2023}. 

Most importantly, we demonstrated that, in the surviving community, there are correlations between the connectivity of a species and its interaction coefficients. These correlations are a fingerprint of the dynamics, which results from constraining a subset of the original species to coexist, and they are not present in the initial pool of species from which the surviving community is formed.

We anticipate that such `patterns' between the network and interaction statistics could be tested in real ecological communities using modern inference techniques \cite{sander2017ecological, maynard2019reconciling}. Perhaps the most generally applicable of our findings is the following `pattern', which we derived using the cavity approach, and which we anticipate ought not to depend much on the specificities of the model in question. In a coexisting community, species with greater abundance will typically have interactions that have a more positive (than the community average) effect on their neighbours' abundances. 

By finding an expression for the degree distribution in the surviving community, we were able to go beyond community wide properties to probe how degree dependent statistics vary across the community. We found that, for a wide range of model parameters, the survival rates and abundance of species are negatively correlated with their degree. This in turn implied that a given species' neighbours were more likely to survive, and were more abundant on average, than said species. We also found that the degree distribution in the surviving community contained relatively few species with high degree, and more species with low degree, than in the original community. This offers a partial mechanism for this same trend found in real ecological networks \cite{williamsSimpleRulesYield2000,camachoAnalyticalSolutionModel2002,camachoRobustPatternsFood2002}, namely that a skewed degree distribution may partially be a consequence of a community's feasibility. 

There are many opportunities for extensions to this work. Our model incorporates only the most simple random network model, but real ecological networks are known to be much more complex. It would be interesting to see how additional structure, such as assortativity, intervality, or particular motifs in the initial network of interactions evolve when we constrain the network to be that of a feasible equilibrium \cite{miloNetworkMotifsSimple2002,stoneNetworkMotifsTheir2019,stoufferQuantitativePatternsStructure2005,stoufferRobustMeasureFood2006,melianFoodWebStructure2002}. We also know that ecological networks are directed networks \cite{garlaschelliPatternsLinkReciprocity2004,garlaschelliUniversalScalingRelations2003}, which would pose a simple mathematical extension to the present work. Finally, ecological networks are known to straddle the line between being dense and sparse, with connectivity widely reported to be in the range $0.05-0.3$ \cite{dunneFoodwebStructureNetwork2002,williamsSimpleRulesYield2000}. Our work could be extended to include sparse corrections for the case where the connectivity is very low (using techniques similar to e.g. Ref. \cite{baronPathIntegralApproach2023}). In particular, we expect a sparse surviving network to have more significant degree correlations than the dense model, which would perhaps offer a mechanism for the disassortativity found in real ecological networks \cite{garlaschelliPatternsLinkReciprocity2004}.

\acknowledgements
We acknowledge support from the Agencia Estatal de Investigaci\'on and Fondo Europeo de Desarrollo Regional (FEDER, UE) under project APASOS (PID2021-122256NB-C21, PID2021-122256NB-C22). This work was partially supported by the María de Maeztu project CEX2021-001164-M funded by the  MICIU/AEI/10.13039/501100011033. We further acknowledge the award of a studentship by the Engineering and Physical Sciences Research Council EPSRC. JWB is supported by grants from the Simons Foundation (\#454935 Giulio Biroli).

\section*{Note}
While this work was being completed we became aware of the preprint \cite{parkIncorporatingHeterogeneousInteractions2024}, in which a very similar model is studied with dynamic mean-field techniques.

\onecolumngrid
\newpage
\appendix
{\begin{center}{\Large -------- Appendix --------}\end{center}}
\setlength{\parskip}{-0.6pt}
\setlength{\parindent}{0pt}
\tableofcontents
\setlength{\parskip}{8pt}
\setlength{\parindent}{0pt}

\let\addcontentsline=\oldaddcontentsline
\let\nocontentsline=\oldaddcontentsline

\section{Derivation of the DMFT effective dynamical equations}
\label{appendix:dmft}
To derive an effective set of mean field equations describing the evolution of species abundances in the community, we start with the MSRJD \cite{De_Dominics1978,martinStatisticalDynamicsClassical1973a,dedominicisFieldtheoryRenormalizationCritical1978a,janssenLagrangeanClassicalField1976b} generating functional of the gLVE dynamics in \cref{eq:glv_dynamcis}
\begin{align}
    Z[\vb* \psi] =& \int \DD{\vb x}\DD{\vb {\wh x}}\exp[i\sum_i\int \dd{t}\wh x_i(t)\qty(\frac{\dot x_i(t)}{x_i(t)} - 1 + x_i(t) - \sum_{j}A_{ij}\alpha_{ij}x_j(t) - h_i(t)) + i\sum_i\int\dd{t}x_i(t)\psi_i(t)], \label{appendix:eq:generating_functional}
\end{align}
where the adjacency matrix $\vb A$ and interaction matrix $\vb* \alpha$ are described in the main text. The functions $h_i(t)$ and $\psi_i(t)$ do not appear in the dynamics in \cref{eq:glv_dynamcis}. These are source fields that are set to zero at the end of the calculation. We will average $Z[\vb*\psi]$ over the distribution of both the network ($\vb A$) and the interactions ($\vb*\alpha$). Like in Ref.~\cite{Galla_2018}, the resulting disorder-averaged functional can then be manipulated into a form that is recognisable as the generating functional of a different, decoupled set of dynamical equations. This set of equations describes the evolution of the abundance of a typical species with degree $k$ in the original community. 

All the disorder in \cref{appendix:eq:generating_functional} is in the term containing the interaction coefficients $A_{ij}\alpha_{ij}$. The average of this term over the distributions of $\vb *\alpha$ and $\vb A$ is carried out as follows
\begin{align}
    &\avg*{\exp[i\sum_{ij}\int\dd{t}A_{ij}\alpha_{ij}\wh x_i(t)x_j(t)]}_{\vb A, \vb*\alpha} \nonumber \\
    &= \prod_{i<j}\avg*{\exp[i\int\dd{t}A_{ij}\qty\Big(\alpha_{ij}\wh x_i(t) x_j(t) + \alpha_{ji}\wh x_j(t)x_i(t))]}_{A_{ij}, (\alpha_{ij}, \alpha_{ji})}\nonumber \\
    &= \prod_{i<j}\qty(1 + \frac{k_ik_j}{dN}\qty(\avg*{\exp[i\int\dd{t}\qty\Big(\alpha\wh x_i(t) x_j(t) + \beta\wh x_j(t)x_i(t))]}_{(\alpha, \beta)} - 1)), \nonumber \\
    &= \exp[\frac{1}{2}\sum_{ij}\ln\qty(1 + \frac{k_ik_j}{dN}\qty(\avg*{\exp[i\int\dd{t}\qty\Big(\alpha\wh x_i(t) x_j(t) + \beta\wh x_j(t)x_i(t))]}_{(\alpha, \beta)} - 1))], \label{appendix:eq:average_Aalpha_ij}
\end{align}
where we have written $\avg{\dots}_{\vb A, \vb*\alpha}$ for a joint average over all elements of the matrices $\vb A$ $\vb*\alpha$, and $\avg{\dots}_{(\alpha_{ij}, \alpha_{ji})}$ for the average over the joint distribution of the specific elements $\alpha_{ij}$ and $\alpha_{ji}$. The joint distribution of $(\alpha_{ij}, \alpha_{ji})$ does not depend on $i$ or $j$, as the pairs $(\alpha_{ij}, \alpha_{ji})$ are drawn independently for each $i$ and $j$. To make this lack of dependence explicit, we have replaced $\alpha_{ij}\to \alpha$ and $\alpha_{ji} \to \beta$ between the second and the third lines.

To decouple the $i$ and $j$ indices from the final expression in \cref{appendix:eq:average_Aalpha_ij}, for each degree $k$ in the original community network, we introduce the following functional
\begin{align}
    P_k[\vb x, \wh{\vb x}] &= \frac{1}{N_k}\sum_{i\in S_k}\prod_t\delta\qty\Big(x(t) - x_i(t))\delta\qty\Big(\wh x(t) - \wh x_i(t)),
\end{align}
where $i \in S_k$ indicates that species $i$ has degree $k$ in the original community and $N_k = p_kN$ is the number of species with degree $k$ in the original community [$p_k$ is the degree distribution in the original community]. For each $k$, the functional $P_k[\vb x, \wh{\vb x}]$ is the probability that the functions $x(t)$ and $\wh x(t)$ are equal to the functions $x_i(t)$ and $\wh x_i(t)$ respectively, which are constrained to follow the dynamics of species with degree $k$. With this definition, we can write the disordered part of the generating functional as 
\begin{align}
    &\avg*{\exp[i\sum_{ij}\int\dd{t}A_{ij}\alpha_{ij}\wh x_i(t)x_j(t)]}_{\vb A, \vb*\alpha} \nonumber \\
    &= \exp\Bigg[\frac{N^2}{2}\sum_{kk'}p_kp_{k'}\int\DD{\vb x}\DD{\wh{\vb x}}\DD{\vb y}\DD{\wh{\vb y}}P_k[\vb x, \wh{\vb x}]P_{k'}[\vb y, \wh{\vb y}]\nonumber \\
    &\qquad \times\ln\qty(1 + \frac{kk'}{dN}\qty(\avg*{\exp[i\int\dd{t}\qty\Big(\alpha\wh x(t) y(t) + \beta\wh y(t)x(t))]}_{(\alpha, \beta)} - 1))\Bigg].
\end{align}
We enforce the definition of $P_k[\vb x, \wh{\vb x}]$ by inserting delta functions in their complex exponential form into the generating functional
\begin{align}
      1 &\propto \int \DD{P_k}\DD{\wh P_k}\exp[i\sum_kN_k\int \DD{\vb x}\DD{\wh{\vb x}} \wh P_k[\vb x, \wh{\vb x}]\qty(P_k[\vb x, \wh{\vb x}] - \frac{1}{N_k}\sum_{i\in S_k}\prod_t\delta\qty\Big(x(t) - x_i(t))\delta\qty\Big(\wh x(t) - \wh f_i(t)))],\nonumber\\
      &\propto \int \DD{P_k}\DD{\wh P_k}\exp[i\sum_kN_k\int \DD{\vb x}\DD{\wh{\vb x}} \wh P_k[\vb x, \wh{\vb x}]P_k[\vb x, \wh{\vb x}] - i\sum_{i\in S_k} \wh P_k[\vb x_i, \wh{\vb x}_i]],
  \end{align}
With these definitions enforced, the disorder averaged generating functional of \cref{eq:glv_dynamcis} takes the following form
\begin{align}
    \avg{Z[\vb* \psi]}_{\vb A, \vb*\alpha} \propto& \int \DD{\vb x}\DD{\wh{\vb x}}\DD{P}\DD{\wh P}\exp[i\sum_i\int \dd{t}\wh x_i(t)\qty(\frac{\dot x_i(t)}{x_i(t)} - 1 + x_i(t) - h_i(t)) + i\sum_i\int\dd{t}x_i(t)\psi_i(t)]\nonumber\\
    &\times\exp\Bigg[\frac{N^2}{2}\sum_{kk'}p_kp_{k'}\int\DD{\vb x}\DD{\wh{\vb x}}\DD{\vb y}\DD{\wh{\vb y}}P_k[\vb x, \wh{\vb x}]P_{k'}[\vb y, \wh{\vb y}]\nonumber \\
    &\qquad\qquad \times\ln\qty(1 + \frac{kk'}{dN}\qty(\avg*{\exp[i\int\dd{t}\qty\Big(\alpha\wh x(t) y(t) + \beta\wh y(t)x(t))]}_{(\alpha, \beta)} - 1))\Bigg] \nonumber \\
    &\times\exp[iN\sum_knp_k\int \DD{\vb x}\DD{\wh{\vb x}} \wh P_k[\vb x, \wh{\vb x}]P_k[\vb x, \wh{\vb x}] - i\sum_{i\in S_k} \wh P_k[\vb x_i, \wh{\vb x}_i]]
\end{align}
To proceed, we could explicitly perform the average over the joint distribution of the interaction strengths $(\alpha, \beta)$ and simplify the resulting expression. If we do this for the interaction statistics in \cref{eq:interaction_statistics}, to leading order in powers of $1/d$, the integrand is of the form $\exp[NS]$. The integral can then be evaluated with a saddle point approximation for large $N$. Even though our interest in the main text is in dense networks, where retaining only leading terms in powers of $1/d$ is valid, we can proceed without having to make this assumption. That is, we can evaluate the integral with a saddle point equation without assuming the network is dense. To do this, we simply assume that the term with a prefactor of $N^2$ in the integrand is $\order{N}$. With this assumption, the following functional is $\order{N^0}$ for each $k$ and $k'$
\begin{align}
    f_{kk'}[\vb x, \wh{\vb x}, \vb y, \wh{\vb y}] = N\ln\qty(1 + \frac{kk'}{dN}\qty(\avg*{\exp[i\int\dd{t}\qty\Big(\alpha\wh x(t) y(t) + \beta\wh y(t)x(t))]}_{(\alpha, \beta)} - 1)) \overset{\text{assumption}}{=} \order{N^{0}},\label{eq:def:f}
\end{align}
we note that if this functional is not $\order{N^0}$, then the following steps in our derivation are not valid. Our expression for the disorder averaged generating functional now reads (the definition of $f_{kk'}$ is just notation, we do not enforce its definition with delta functions)
\begin{align}
    \avg{Z[\vb*\psi]}_{\vb A, \vb*\alpha} \propto& \int \DD{P}\DD{\wh P}\nonumber\\
    &\times\exp[\frac{N}{2}\sum_{kk'}p_kp_{k'}\int\DD{\vb x}\DD{\wh{\vb x}}\DD{\vb y}\DD{\wh{\vb y}}P_k[\vb x, \wh{\vb x}]P_{k'}[\vb y, \wh{\vb y}]f_{kk'}[\vb x, \wh{\vb x}, \vb y, \wh{\vb y}]]\nonumber \\
    &\times\exp[iN\sum_kp_k\int \DD{\vb x}\DD{\wh{\vb x}} \wh P_k[\vb x, \wh{\vb x}]P_k[\vb x, \wh{\vb x}]]\nonumber \\
    &\times\exp[N\sum_kp_k\ln\int\DD{\vb x}\DD{\wh{\vb x}}\exp[i\int \dd{t}\wh x(t)\qty(\frac{\dot x(t)}{x(t)} - 1 + x(t) - h_k(t)) - i\wh P_k[\vb x, \wh{\vb x}] + i\int\dd{t}x(t)\psi_k(t)]], \label{appendix:eq:disorder_averaged_Z}
\end{align}
where we have supposed that $h_i(t)$ and $\psi_i(t)$ only depend on the degree of species $i$ in order to be able to factorise the final term. That is, $h_i(t) = h_k(t)$ and $\psi_i(t) = \psi_k(t)$ for all species $i$ with degree $k$. We can evaluate this integral with a saddle-point approximation for large $N$. First, taking the derivative of the exponent with respect to the hatted functionals $\wh P_k$ gives
\begin{align}
    P_k[\vb x, \wh{\vb x}] &= \frac{\exp[i\int \dd{t}\wh x_k(t)\qty(\frac{\dot x_k(t)}{x_k(t)} - 1 + x_k(t) - h_k(t)) - i\wh P_k[\vb x_k, \wh{\vb x}_k] + i\int\dd{t}x_k(t)\psi_k(t)]}{\int \DD{\vb x}\DD{\wh{\vb x}}\exp[i\int \dd{t}\wh x(t)\qty(\frac{\dot x(t)}{x(t)} - 1 + x(t) - h_k(t)) - i\wh P_k[\vb x, \wh{\vb x}] + i\int\dd{t}x(t)\psi_k(t)]}, \label{appendix:eq:hatted_saddle}
\end{align}
where the subscript in $x_k(t)$ indicates that the abundance $x_k(t)$ is constrained to be the trajectory of a typical species with degree $k$, we justify this interpretation further on in the derivation. The unhatted saddle equation (taking derivatives of the exponent with respect to $P_k$) is
\begin{align}
    \wh P_k[\vb x, \wh{\vb x}] &= i\sum_{k'}p_{k'}\int\DD{\vb y}\DD{\wh{\vb y}}P_{k'}[\vb y, \wh{\vb y}]f_{kk'}[\vb x, \wh{\vb x}, \vb y, \wh{\vb y}].\label{appendix:eq:unhatted_saddle} 
\end{align}
Substituting \cref{appendix:eq:hatted_saddle} into \cref{appendix:eq:unhatted_saddle}, we arrive at the following self-consistent functional equation for $\wh P_k$ 
\begin{align}
    \wh P_k[\vb x, \wh{\vb x}] =  i\sum_{k'}p_{k'}\frac{\int\DD{\vb y}\DD{\wh{\vb y}}~f_{kk'}[\vb x, \wh{\vb x}, \vb y, \wh{\vb y}]~\exp[i\int \dd{t}\wh y(t)\qty(\frac{\dot y(t)}{y(t)} - 1 + y(t) - h_{k'}(t)) - i\wh P_{k'}[y, \wh y] + i\int\dd{t}y(t)\psi_{k'}(t)]}{\int \DD{\vb y}\DD{\wh{\vb y}}\exp[i\int \dd{t}\wh x(t)\qty(\frac{\dot y(t)}{y(t)} - 1 + y(t) - h_{k'}(t)) - i\wh P_{k'}[y, \wh y] + i\int\dd{t}y(t)\psi_{k'}(t)]}. \label{appendix:eq:effective_disorder_average}
\end{align}
We now re-write this as 
\begin{align}
    \wh P_k[\vb x, \wh{\vb x}] =  i\sum_{k'}p_{k'}\avg*{f_{kk'}[\vb x, \wh{\vb x}, \vb y, \wh{\vb y}]}^{(y)}_{k'}.    
\end{align}
where $\avg{\dots}^{(y)}_{k'}$ stands in for the ratio of functional integrals in \cref{appendix:eq:effective_disorder_average}, with $(\dots)$ in place of $f_{kk'}[\vb x, \wh{\vb x}, \vb y, \wh{\vb y}]$. In particular, we point out that this means the quantity $\avg{f_{kk'}[\vb x, \wh{\vb x}, \vb y, \wh{\vb y}]}_y$ does not depend on arguments $\vb y$, $\wh{\vb y}$ or $y$, but it \textit{does} depend on the degree $k$. To interpret $\avg{\dots}^{(y)}_k$, we compare functional derivatives of the expression for $Z$ in \cref{appendix:eq:generating_functional,appendix:eq:disorder_averaged_Z} with respect to $\psi_i(t)$ and $h_i(t)$ to find (in the limit of large $N$)
\begin{align}
    \avg{x(t)}_k &= -i\frac{1}{N_k}\sum_{i\in S_k}\eval{\pdv{\avg{Z[\vb*\psi]}_{\vb A, \vb*\alpha}}{\psi_i(t)}}_{h=\psi=0} = \frac{1}{N_k}\sum_{i\in S_k}\avg{x_i(t)}_{\vb A, \vb*\alpha}, \\
    \avg{\wh x(t)}_k &= i\frac{1}{N_k}\sum_{i\in S_k}\eval{\pdv{\avg{Z[\vb*\psi]}_{\vb A, \vb*\alpha}}{h_i(t)}}_{h=\psi=0} = \frac{1}{N_k}\sum_{i\in S_k}\avg{\wh x_i(t)}_{\vb A, \vb*\alpha}, \\
    \avg{\wh x(t) x(t')}_k &= -\frac{1}{N_k}\sum_{i\in S_k}\eval{\pdv{\avg{Z[\vb*\psi]}_{\vb A, \vb*\alpha}}{h_i(t)}{\psi_i(t')}}_{h=\psi=0} = \frac{1}{N_k}\sum_{i\in S_k}\avg{\wh x_i(t)x_i(t')}_{\vb A, \vb*\alpha},
\end{align}
where we have dropped the superscripts $(x)$ in e.g. $\avg{x(t)}_k^{(x)}$ on the LHS of the above equations as there is only one dynamical variable which could be averaged over. We can do the same calculation for any other powers of $x(t)$ and $\wh x(t)$. Hence, $\avg{x(t)}_k$ is, in the large $N$ limit, equal to the average abundance of species with degree $k$ in the community. Further, because $Z[\vb*\psi=0]=1$ (\cref{appendix:eq:generating_functional} is the integral of a delta function when $\psi_i(t) = 0$), derivatives of $Z$ with respect to factors of $\vb h$ only are all zero. Therefore, any averages containing only hatted variables vanish.

At the saddle point, we can finally write the disorder averaged generating functional as 
\begin{align}
    \avg{Z[\vb* \psi]}_{\vb A, \vb*\alpha} \propto &\int\DD{\vb x}\DD{\wh{\vb x}}\exp[i\sum_kp_k\int \dd{t}\wh x_k(t)\qty(\frac{\dot x_k(t)}{x_k(t)} - 1 + x_k(t) - h_k(t))]\nonumber \\
    &\times\exp[\sum_{kk'}p_kp_{k'}\avg{f_{kk'}[\vb x_k, \wh{\vb x}_k, \vb y, \wh{\vb y}]}^{(y)}_{k'} + i\sum_kp_k\int\dd{t}x_k(t)\psi_k(t)].\label{appendix:eq:Z_final}
\end{align}

We now evaluate the functional $f_{kk'}[\vb x, \wh{\vb x}, \vb y, \wh{\vb y}]$, with the random matrix $\vb*\alpha$ as described in \cref{section:model} of the main text. To leading order in $1/N$, we find
\begin{align}
    f_{kk'}[\vb x_k, \wh{\vb x}_k, \vb y, \wh{\vb y}] =& i\frac{kk'\mu}{d^2}\int\dd{t}\qty\Big(\wh x_k(t) y(t) + \wh y(t) x_k(t))\nonumber\\
    & - \frac{kk'\sigma^2}{2d^2}\qty[\qty(\int\dd{t}\wh x_k(t) y(t))^2 + \qty(\int\dd{t}\wh y(t) x_k(t))^2] - \frac{kk'\gamma\sigma^2}{d^2}\int\dd{t}\dd{t'}\wh x_k(t) y(t)\wh y(t') x_k(t').
\end{align}
Averaging over the dynamics of $y$, we are left with 
\begin{align}
    \avg{f_{kk'}[\vb x_k, \wh{\vb x}_k, \vb y, \wh{\vb y}]}^{(y)}_{k'} =& i\frac{kk'\mu}{d^2} \int \dd{t}\wh x_k(t)\avg{y(t)}_{k'} - \frac{kk'\sigma^2}{2d^2}\int\dd{t}\dd{t'}\qty\Big(\wh x_k(t)\wh x_k(t')\avg{y(t)y(t')}_{k'} + 2\gamma\wh x_k(t)x_k(t')\avg{y(t)\wh y(t')}_{k'}).
\end{align}
Substituting this into \cref{appendix:eq:Z_final}, we recognise $\avg{Z[\vb* \psi]}_{\vb A, \vb*\alpha}$ as the generating functional of the following set of effective dynamical equations 
\begin{align}
      \dot x_k(t) = x_k(t)\qty(1 - x_k(t) + \frac{\mu k}{d^2} \sum_{k'} p_{k'}k'M_{k'}(t) + \frac{\gamma\sigma^2k}{d^2}\sum_{k'} p_{k'}k'\int\dd{t}G_{k'}(t, t')x_k(t') + \eta_k(t)), \label{appendix:eq:effective_dynamics}
\end{align}
where we have set $h_k(t) = 0$ and written $\avg{y(t)}_k = M_k(t)$ and $-i\avg{\wh y(t')y(t)}_k = G_k(t, t')$. The quantities $M_k(t), G_k(t, t')$, and the colored gaussian noise term $\eta_k(t)$ are determined self-consistently via the following equations 
\begin{align}
    \avg{\eta_k(t)}_\eta &= 0, \nonumber\\
    \avg{\eta_k(t)\eta_{l}(t')}_\eta &= \delta_{kl}\frac{\sigma^2k}{d^2}\sum_{k'} p_{k'}k'\avg{x_k(t)x_{k'}(t')}_\eta, \nonumber\\
    M_k(t) &= \avg{x_k(t)}_\eta, \nonumber\\
    G_k(t, t') &= \fdv{\avg{x_k(t)}_\eta}{\eta_k(t')}.\label{appendix:eq:effective_dynamics_averages}
\end{align}
The last of these relationships follows from writing down the generating funcitonal of the effective dynamics \cref{appendix:eq:effective_dynamics} without performing the average over the noise term $\eta_k(t)$ (which would simply return \cref{appendix:eq:Z_final}). From this functional, it is then clear that differentiation with respect to the noise term `pulls down' a factor of $-i\wh x_k(t)$, hence we can replace factors of $-i\wh x(t)$ in \cref{appendix:eq:Z_final} with derivatives with respect to the noise.
\subsection{Fixed point equations}\label{appendix:fixed_point_equations}
As discussed in the main text, we can derive a closed set of self-consistent equations for the abundance distribution in the surviving community at a fixed point of the dynamics. Suppose that the dynamics in \cref{appendix:eq:effective_dynamics} reaches a fixed point $x^\star_k$. In this case, the noise term will be a static, mean zero guassian random variable with variance $\sigma^2k/d^2\sum_{k'}p_{k'}k'q_k$, with $q_k = \avg{(x_k^\star)^2}_z$. We also assume that the system's response function is a function of time differences only in this regime, so that $G_k(t, t') = G_k(t - t')$, which means we can write $\int\dd{t'}G_k(t, t')x_k(t') = \chi_k x^\star_k$ for $\chi_k = \int \dd{\tau} g_k(\tau)$. The fixed point abundance distribution of $x_k^\star$ is then given by (we have added back in the factor of $h$ from the original generating functional so that we can cleanly write down the definition of $\chi_k$)
\begin{align}
    x_k^\star(z) = \max\qty(0, \frac{1 + \mu k\sum_{k'}p_{k'}k'M_{k'}/d^2 + z \sigma\sqrt{k\sum_{k'}p_{k'}k'q_{k'}}/d + h}{1 - \gamma\sigma^2k\sum_{k'}p_{k'}k'\chi_{k'}/d^2}).\label{appendix:eq:fixed_point_abundance}
\end{align}
By carefully evaluating the definitions of the parameters $M_k, q_k$, and $\chi_k$, we arrive at the fixed point equations in the main text, which we repeat here
\begin{align}
      M_k &= \int_{x_k^\star>0} \dd{z} P(z) x^\star_k(z), \nonumber \\
      q_k &= \int_{x_k^\star>0} \dd{z} P(z) x^\star_k(z)^2, \nonumber \\
      \chi_k &= \int_{x_k^\star>0} \dd{z} P(z) \pdv{x^\star_k(z)}{h}, \label{appendix:eq:fixed_point_equations}
\end{align}
where $P(z) = e^{-z^2/2}/\sqrt{2\pi}$ is the probability density function of the standard normal distribution. The survival probability for species with degree $k$ in the original community is also determined from the fixed point abundances
\begin{align}
    \phi_k = \int_{x^\star_k>0}\dd{z} P(z).
\end{align}

We now expand out the definitions in \cref{appendix:eq:fixed_point_equations} to find the explicit set of equations which we can numerically solve. Expanding the definitions gives (note that the integration region $x^\star_k>0$ needs to be converted into an integration region over $z$ using \cref{appendix:eq:fixed_point_abundance}) 
\begin{align}
      \chi_k &= \frac{w_0(\Delta_{k})}{1 - \frac{\gamma\sigma^2k}{d^2}\sum_{k'}p_{k'}k'\chi_{k'}}, \nonumber\\
      M_k &= \Sigma_k w_1(\Delta_k),\nonumber \\
      q_k &= \Sigma^2_kw_2(\Delta_k),\label{appendix:eq:fixed_point_explicit}
\end{align}
where we have defined the shorthands
\begin{align}
    \Sigma_k &= \frac{\sigma\sqrt{k\sum_{k'}p_{k'}k'q_{k'}}/d}{1 - \gamma\sigma^2k\sum_{k'}p_{k'}k'\chi_{k'}/d^2}, \nonumber\\
    \Delta_k &= \frac{1 + \mu k\sum_{k'}p_{k'}k'M_{k'}/d^2}{\sigma\sqrt{k\sum_{k'}p_{k'}k'q_{k'}}/d},\nonumber\\
    w_l(\Delta_k) &= \int_{-\infty}^{\Delta_k} \dd{z}P(z)\qty(\Delta_k - z)^l.\label[plural_equation]{appendix:eq:fixed_point_abbreviations}
\end{align}
We also define the probability of survival for species that have degree $k$ in the original community
\begin{align}
      \phi_k = \int_{x_k^\star>0}\dd{z} P(z),
\end{align}
as well as the community wide abundance and survival probability 
\begin{align}
    M = \sum_kp_kM_k,\qquad \phi = \sum_kp_k\phi_k.
\end{align}

\cref{appendix:eq:fixed_point_explicit} can be numerically solved to find the fixed point parameters for specific degree $k$. The integrals defining $w_l(\Delta_k)$ can be explicitly evaluated. For $l = 0, 1, 2$ we have 
\begin{align}
    w_0(x) &= \frac{1}{2}\qty(1 + \erf\qty(\frac{x}{\sqrt{2}})), \nonumber\\
    w_1(x) &= P(x) + \frac{1}{2}x\qty(1 + \erf\qty(\frac{x}{\sqrt{2}})), \nonumber\\
    w_2(x) &= xP(x) + \frac{1}{2}\qty(1 + x^2)\qty(1 + \erf\qty(\frac{x}{\sqrt{2}}))
\end{align}

Practically, we solve the fixed point equations using the function `scipy.optimize.root' in python, with an initial guess $\{\chi_k\}, \{M_k\}, \{q_k\}$ either determined by a previously found solution with similar values of the parameters $\mu, \sigma,\gamma, p_k$, or by running the GLV dynamics themselves for small $N$ about $50$ times to obtain empirical estimates for $\phi_k, M_k$, and $q_k$. We then use the relation (derived from the fixed point equations) $\chi_k = M_k\phi_k w_1(w_0^{-1}(\phi_k))/\sqrt{\sigma^2k\sum_{k'}p_{k'}k'q_{k'}/d^2}$ to determine a sensible initial guess for $\chi_k$. Here $w_0^{-1}$ is the inverse function of $w_0$.
\subsection{The abundance distribution}\label{appendix:abundance_distribution}
The abundance distribution $\text{AD}_k(x)$ for species with degree $k$ in the original community is derived from \cref{appendix:eq:fixed_point_abundance}. It has the general form $\text{AD}_k(x) = (1 - \phi_k)\delta(x) + \Theta(x)P_{m_k, \Sigma_k}(x)$, where $\Theta(x) = 1$ if $x > 0$ and is zero otherwise. $P_{m_k, \Sigma_k}(x)$ is a gaussian PDF with mean $m_k$ and variance $\Sigma_k^2$, $\Sigma_k$ is defined in \cref{appendix:eq:fixed_point_abbreviations}, and $m_k$ is given by the following expression
\begin{align}
    m_k = \frac{1 + \mu k\sum_{k'}p_{k'}k'M_{k'}/d^2}{1 - \gamma\sigma^2k\sum_{k'}p_{k'}k'\chi_{k'}/d^2}.
\end{align}
The community-wide abundance distribution $\text{AD}(x)$, such as the one plotted in \cref{fig:abundance_distribution} in the main text, is equal to the weighted average of the individual degree distributions $\text{AD}(x) = \sum_kp_k~\text{AD}_k(x)$.

\section{Trend of \texorpdfstring{$\phi_k$ and $M_{k^\star}$}{} with degree}\label{appendix:phik_Mk_decreasing_functions}
In the main text, we claim that, for $\mu < 0$, the survival rate $\phi_k$ is always a decreasing function of the degree $k$, and that the same is true for a wide range of parameters for $M_k$. In this section we justify these claims.
\subsection{Trend of \texorpdfstring{$\phi_k$ with $k$}{}}
By definition, we can express the survival probability for species with degree $k$ in the original community as [see \cref{appendix:eq:fixed_point_explicit}]
\begin{align}
    \phi_k = w_0(\Delta_k),
\end{align}
where $\Delta_k$ is defined in \cref{appendix:eq:fixed_point_abbreviations}. The function $w_0$ is an increasing function of its argument. Hence, $\phi_k$ is an increasing(decreasing) function of $k$ precisely when $\Delta_k$ is an increasing(decreasing) function of $k$. For fixed model parameters, $\Delta_k$ has the following functional dependence on the degree $k$ 
\begin{align}
    \Delta_k = \frac{1}{S}\qty(\frac{1}{\sqrt{k}} + U\sqrt{k}), \label{appendix:eq:delta_k_general_form}
\end{align}
where $S$ and $U$ do not depend on $k$ explicitly and are given by 
\begin{align}
    S &= \frac{\sigma}{d}\sqrt{\sum_{k'}p_{k'}k'q_{k'}}, \nonumber \\
    U &= \frac{\mu}{d^2}\sum_{k'}p_{k'}k'M_{k'}. \label[plural_equation]{appendix:eq:U_and_S}
\end{align}
Differentiating \cref{appendix:eq:delta_k_general_form} with respect to $k$, we find that $\Delta_k$ is stationary in $k$ when 
\begin{align}
    k = \frac{1}{U}.\label{appendix:eq:phi_k_stationary}
\end{align}
If $\mu < 0$, then $U < 0$ also, and the LHS and RHS have opposite signs (all other components in the equation are positive by definition), so there is no stationary point. Hence, $\phi_k$ is a decreasing function of $k$. If, on the other hand, $\mu$ is positive, then $\phi_k$ is decreasing provided the LHS is smaller then the RHS, if the LHS is larger, then the trend reverses. This becomes increasingly likely for more positive $\mu$ and larger abundances $M_k$.
\begin{figure}
    \centering
    \includegraphics[width=0.5\textwidth]{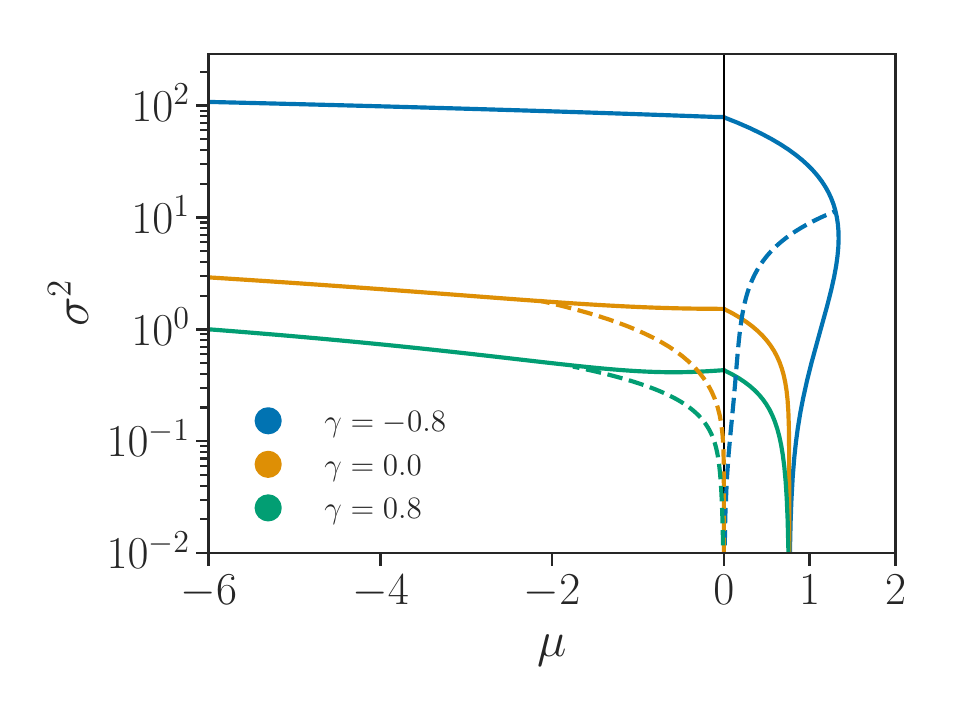}
    \caption{Curves for which $M_{k^\star}$ and $k^\star$ are uncorrelated (dashed lines). To the left of the dashed lines, $M_{k^\star}$ and $k^\star$ are negatively correlated, and to the right of the dashed lines $M_{k^\star}$ and $k^\star$ are positively correlated. The area underneath the solid curves is stable, and the area above is unstable (see \cref{fig:stability} in the main text for more detail).}
    \label{appendix:fig:Mk_uncorrelated}
\end{figure}
\subsection{Trend of \texorpdfstring{$M_{k^\star}$ with $k^\star$}{}}
We can follow the same procedure as for $\phi_k$ to find the degree at which the abundance $M_k$ has a stationary point. It is stationary when 
\begin{align}
    k = \frac{\phi_k - M_k}{U\phi_k + TM_k},
\end{align}
where 
\begin{align}
    T = \frac{\gamma\sigma^2}{d^2}\sum_{k'}p_{k'}k'\chi_{k'}.
\end{align}
This condition is not as straightforward to analyse as the equivalent condition for the stationary point of $\phi_k$ in \cref{appendix:eq:phi_k_stationary}. However, we can still find the general trend of $M_{k^\star}$ with $k^\star$ by noting (as we do in the main text) that the covariance $\covariance(M_{k^\star}, k^\star) = \qty(\sum_{k^\star}p_{k^\star} M_{k^\star} k^\star) - \qty(\sum_{k^\star}p_{k^\star}M_{k^\star})\qty(\sum_{k^\star}p_{k^\star} k^\star)$ is negative whenever $M_{k^\star}$ is a decreasing function of $k^\star$. In \cref{appendix:fig:Mk_uncorrelated}, we plot the curve satisfying $\covariance(M_{k^\star}, k^\star) = 0$ in the $(\mu, \sigma^2)$ plane for different values of $\gamma$ and fixed network structure. Demonstrating that when $\mu < 0$, only a small range of parameters in the stable phase give rise to positive correlations (right of the curve), hence our focus on this trend in particular in the main text. 
\section{Stability}\label{appendix:stability}
\subsection{Diverging abundances}
To numerically find the boundary between the fixed point and the point at which the average abundance $M$ diverges, we first express the fixed point equations in \cref{appendix:eq:fixed_point_explicit} in terms of the new variables $\chi_k, \tilde M_k = M_k/\sqrt{\sum_{k'}p_{k'}q_{k'}}$, and $\tilde q_k = q_k/\sum_{k'}p_{k'}q_{k'}$. Unlike $M_k$ and $q_k$, $\tilde M_k$ and $\tilde q_k$ remain finite when the average abundance diverges. In the limit of infinite average abundance, the new fixed point equations are equivalent to \cref{appendix:eq:fixed_point_explicit} with the replacements $M_k \to \tilde M_k, q_k \to \tilde q$, and $\Delta_k \to \tilde \Delta_k$, where 
\begin{align}
    \tilde \Delta_k &= \frac{\mu k\sum_{k'}p_{k'}k'\tilde M_{k'}/d^2}{\sigma\sqrt{k\sum_{k'}p_{k'}k'\tilde q_{k'}}/d}.
\end{align}
Solving these new fixed point equations, together with the condition $\sum_k \tilde q_k = 1$, gives the diverging abundance curves in \cref{fig:stability} in the main text.
\subsection{Linear instability}
To derive the stability condition \cref{eq:linear_stability} from the main text, we consider a linear perturbation to the effective dynamics \cref{appendix:eq:effective_dynamics} near a fixed point. We follow along the lines of the stability analyses in \cite{1overf_noise,Galla_2018} (see also \cite{poleyGeneralizedLotkaVolterraModel2023} for a derivation involving block structured interactions, which generalises the present argument).

The local stability of possible fixed points can be probed by addition of an infinitesimal independent and identically distributed Gaussian perturbation $\epsilon \xi_k(t)$ to each equation in the effective dynamics \cref{appendix:eq:effective_dynamics}. In the stable regime, we expect the system to return to the fixed point when perturbed, that is, we expect the response of the system to the perturbation to decay to zero as $t\to \infty$.

Adding the perturbation $\epsilon \xi_k(t)$ to the effective dynamics [\cref{appendix:eq:effective_dynamics}], we have 
\begin{align}
      \dot x_k(t) = x_k(t)\qty(1 - x_k(t) + \frac{\mu k}{d^2} \sum_{k'} p_{k'}k'M_{k'}(t) + \frac{\gamma\sigma^2k}{d^2}\sum_{k'} p_{k'}k'\int\dd{t}G_{k'}(t, t')x_k(t') + \eta_k(t) + \epsilon\xi_k(t)), \label{appendix:eq:perturbed_effective_dynamics}
\end{align}
where $M_k(t), G_k(t)$, and the noise term $\eta_k(t)$ are defined in \cref{appendix:eq:effective_dynamics_averages}.

We quantify the linear response of $x_k(t)$ and $\eta_k(t)$ to the perturbation $\epsilon\xi(t)$ about the fixed point by $y_k(t)$ and $\zeta_k(t)$ respectively, so that
\begin{align}
	x_k(t) &= x^\star_k + \epsilon y_k(t) \nonumber \\
	\eta_k(t) &= \eta^\star_k + \epsilon \zeta_k(t).
\end{align}
From this, we can self-consistently relate the responses to each other using \cref{appendix:eq:effective_dynamics_averages}
\begin{align}
\avg{\zeta^a(t) \zeta^a(t')} &= \frac{\sigma^2k}{d^2}\sum_{k'} p_{k'}k' \avg{y_k(t) y_k(t')}.
\end{align}
Assuming time translation invariance of the fixed point in the long-time limit, we obtain the following equation for the evolution of the perturbed abundances 
\begin{align}
	\dot{y}_k(t) = x_k^\star \qty(- y_k(t) + \frac{\gamma\sigma^2k}{d^2}\sum_{k'}p_{k'}k' \int_0^t dt'~G_k(t - t') y_k(t')  + \zeta_k(t) + \xi_k(t)). \label{app:eq:linAroundNonZeroFP}
\end{align}

We now follow \cite{1overf_noise,Galla_2018} by taking the Fourier transform (denoted with a hat, which we note is not related to the hatted variables in the generating functional calculations in \cref{appendix:dmft})
\begin{align}
    i\omega \hat{y}_k(\omega) = x_k^\star\qty(-\hat y_k(\omega) + \frac{\gamma\sigma^2k}{d^2}\sum_{k'}p_{k'}k'\hat G_k(\omega)\hat y_k(\omega) + \hat \zeta_k(\omega) + \hat \xi_k(\omega)).
\end{align}
Squaring and averaging over the distribution of the perturbing noise $\xi_k$
\begin{align}
    \qty{\frac{|\omega|^2}{(x^\star_k)^2} + \Big|1 - \frac{\gamma\sigma^2k}{d^2}\sum_{k'}p_{k'}k'\hat G_k(\omega)\Big|^2} \avg{|\hat y_k(\omega)|^2}_\xi  = \phi_k \qty{\frac{\sigma^2k}{d^2}\sum_{k'}p_{k'}k' \avg{|\hat y_{k'}(\omega)|^2}_\xi + 1},
\end{align}
where the factor of the survival rate $\phi_k$ is due to the fact that \cref{app:eq:linAroundNonZeroFP} only applies to non-zero fixed points. Fluctuations around the zero point decay and hence do not contribute to $\avg{|\hat y_k(\omega)|^2}$. Noticing that $\hat G_k(0) = \chi^a$, we now set $\omega = 0$ (see \cite{1overf_noise}) and find
\begin{align}
    \qty{1 - \frac{\gamma\sigma^2k}{d^2}\sum_{k'}p_{k'}k'\chi_{k'}}^2 Y_k &= \phi_k\qty{\frac{\sigma^2k}{d^2}\sum_{k'}p_{k'}k' Y_{k'} + 1}, \label{app:eq:StabilityGeneralCond}
\end{align}
where $Y_k \equiv \avg{|\hat y_k(0)|^2}$. Assuming a stationary state in which $\avg{y_k(t)y_k(t + \tau)}$ depends on $\tau$ only, then $Y_k = \int \dd{\tau}\avg{y_k(t)y_k(t + \tau)}$. In the stable regime, $y_k(t)\to 0$ as $t\to 0$, and therefore $Y_k$ is finite. Hence, if $Y_k$ is not finite, then this signals the onset of linear instability. Hence, the onset of linear instability corresponds to the point at which the only solution to \cref{app:eq:StabilityGeneralCond} for which $Y_k$ diverges for some degree $k$.

\cref{app:eq:StabilityGeneralCond} is equivalent to the condition for linear (in)stability given in the main text [\cref{eq:linear_stability}]. To make the connection, we first use the fixed point equations (\ref{appendix:eq:fixed_point_explicit}) (the one for $\chi_k$) to re-arrange \cref{app:eq:StabilityGeneralCond} into the following form 
\begin{align}
      \tilde Y_k = \frac{\sigma^2k\chi_k^2}{d^2\phi_k}\sum_{k'}p_{k'}\qty[k' + \frac{1}{\sum_{k''}p_{k''}Y_{k''}}]\tilde Y_{k'}, \label{appendix:eq:StabilityGeneralCondEigval}
\end{align}
where we have defined $\tilde Y_k = Y_k/(\sum_{k'}p_{k'} Y_{k'})$, which remains finite, even if $Y_k$ diverges for some $k$. At the point of linear instability, the quantity $1//(\sum_{k'}p_{k'} Y_{k'}) = 0$, and \cref{appendix:eq:StabilityGeneralCondEigval} is an eigenvalue equation 
\begin{align}
    Y_k = \sum_{k'} S_{kk'} Y_{k'},\label{appendix:eq:linear_stability_eigenvalue_equation}
\end{align}
where $\vb S$ is a matrix with $kk'$ element equal to $S_{kk'} = \sigma^2k\chi_k^2p_{k'}k'/(d^2\phi_k)$. All elements of the matrix $\vb S$ are positive, as are all elements of the vector $\vb Y$ (by definition of $Y_k$). Hence, $\vb Y$ is the Perron-Frobenius (PF) eigenvector of $\vb S$, with PF eigenvalue equal to one. Further, as $\vb S$ can be written as the outer product of two vectors ($\vb S = \vb a \vb b^T$, for $a_k = \sigma^2k\chi_k^2/(d^2\phi_k)$ and $b_k = p_{k}k$), its PF eigenvalue is given by the \textit{inner} product of these two vectors ($\vb a^T \vb b = \sum_k a_kb_k$). That is, the linear instability occurs at the point when 
\begin{align}
    \frac{\sigma^2}{d^2}\sum_kp_k\frac{k^2\chi_k^2}{\phi_k} = 1.\label{appendix:eq:linear_instability}
\end{align}
To see that the LHS is smaller than one in the stable regime, we repeat the same argument, but we write $Y = \sum_kp_kY_k$ in \cref{appendix:eq:StabilityGeneralCondEigval} and do not assume that $Y$ diverges. This time, the eigenvalue condition is different, we have $\lambda_{\text{PF}}[\vb S + \vb p/Y] = 1$, where we have defined the matrix $\vb p$ with elements $p_{kk'} = \sigma^2\chi_k^2p_{k'}/(d^2\phi_k)$. All elements of the matrix $\vb p$ are positive. It is known that the PF eigenvalue of a matrix is an increasing function of its elements \cite{cohenDerivativesSpectralRadius1978,deutschDerivativesPerronRoot1984}. Hence, we have $\lambda_{\text{PF}}[\vb S] \leq \lambda_{\text{PF}}[\vb S + \vb p/Y] = 1$ in the stable regime. That is, in the stable regime, the LHS of \cref{appendix:eq:linear_instability} is less than one.

In \cref{fig:stability} in the main text, we solve for the boundary of the linearly stable region in parameter space by adding the condition in \cref{appendix:eq:linear_instability} to the fixed point equations.
\section{Interaction statistics in the surviving community}
\label{appendix:interaction_statistics}
To find the quantities defined in \cref{eq:muinmuout}, we employ a similar philosophy to that was used in Ref. \cite{baronBreakdownRandomMatrixUniversality2023b}, except here we use a cavity approach. In this case, the cavity approach helps to elucidate the mechanism behind the trends seen in \cref{fig:interaction_degree_correlations} by revealing the origins of all the contributing factors to the quantities in \cref{eq:muinmuout}.

\subsection{The cavity approach}
At the fixed point, we must have that
\begin{align}
      x_i = \max\qty(0, 1 + \sum_{j} A_{ij} \alpha_{ij} x_j + h_i), \label{eq:statvalue}
\end{align}
where $h_i$ is again an external field that we include for analytical purposes, and which we later set to zero.

Let us suppose that we introduce a new `cavity' species, which we endow with index $0$, to the network. We suppose that the cavity species has degree $k$ (in the original community). Let us now inspect the following quantity
\begin{align}
    \mu_0^{(k)} = \sum_{j} A_{0j} \alpha_{0j} \theta_j, \label{eq:mu0def}
\end{align}
where $\theta_j(x_j) =0$ if a species is extinct and $\theta_j(x_j) =1$ if the species survives. Following the usual procedure for cavity calculations, we attempt to find $\theta_j$ in terms of system parameters \textit{before} the addition of the cavity species. We write
\begin{align}
    \theta_j = \theta_j^{(0)} + \delta\theta_j  ,
\end{align}
so that $\delta\theta_j$ (which takes values $0$ or $\pm1$) accounts for the changes in the numbers of surviving species due to the introduction of the cavity species. We expect the number of species that go extinct due to the introduction of the new species to be small.

Inserting this into \cref{eq:mu0def} and defining $\alpha_{ij} = \mu/d + a_{ij}$, we find 
\begin{align}
    \mu_0^{(k)} = \frac{ \mu}{d} \sum_{j} A_{0j}\theta_j + \sum_{j} A_{0j} a_{0j} \theta^{(0)}_j + \sum_{j}   A_{0j} \alpha_{0j} \delta\theta_j. \label{eq:cavity1}
\end{align}
Let us now use the fact that we have many species to write each of these terms in terms of the statistics of the community. 

\subsection{Averaging \texorpdfstring{\cref{eq:cavity1}}{} over the wider pool of species}

Let us now discuss the statistics of each of the terms in \cref{eq:cavity1}, keeping the cavity interactions $A_{0j}$, $\alpha_{0j}$ and $\alpha_{j0}$ fixed. Specifically, we first take the mean with respect to the interactions $\alpha_{ij}$ and the network $A_{ij}$, where both $i$ and $j$ are not equal to $0$, and then discuss the variance with respect to these same quantities. We denote the average with respect to these variables as $\langle \cdot \rangle_0$ (as opposed to $\langle \cdot\rangle$, which indicates an average over all interaction coefficients, including those of the cavity species). 

We treat the average over the random variables $A_{0j}$, $\alpha_{0j}$, $A_{j0}$, and $\alpha_{j0}$ separately so that we can better see how $\mu_0^{(k)}$ relates to other random quantities of interest, for example the abundance $x_0$. This will help us to understand the origin of the behaviour of $\mu_k^{(0)}$ as $k$ is varied (shown in \cref{fig:interaction_degree_correlations}).

Taking for example the first term in \cref{eq:cavity1}, we first examine its mean, and then its fluctuations. We find that 
\begin{align}
    \frac{ \mu}{d}\left\langle \sum_{j} A_{0j}\theta_j \right\rangle_0 =  \frac{ \mu}{d}\sum_j A_{0j} \left[\left\langle \theta_j^{(0)}\right\rangle_0 +   \langle \delta\theta_j \rangle_0\right],
\end{align}
where we have used that the survival of species before the introduction of the cavity is independent of $A_{0j}$. We can rewrite the second of these terms using the fact that \begin{align}
    \langle\delta \theta_j\rangle_0 = \frac{d\phi_{k_j}}{dh} A_{j0}\alpha_{j0}x_0, \label{eq:deltaphi}
\end{align} 
where we write $k_j$ for the degree of species $j$, and we obtain
\begin{align}
    \frac{ \mu}{d}\left\langle \sum_{j} A_{0j}\theta_j \right\rangle_0 =  \frac{ \mu}{d}\sum_{j} A_{0j} \left[\phi_{k_j} +   \frac{d\phi_{k_j}}{dh} \alpha_{j0}x_0\right].
\end{align}
We see that the second of these terms is a small $O(1/N)$ correction compared to the first, and hence we can ignore it.  We now average over both the interaction statistics of the original community and the cavity species to obtain
\begin{align}
    \frac{ \mu}{d}\left\langle \sum_{j} A_{0j}\theta_j \right\rangle \approx \frac{ \mu}{d}\sum_{k'} \frac{kk'}{dN} \sum_{j \in S_{k'}} \left\langle \theta_j^{(0)}\right\rangle_0= \frac{k}{d^2}\sum_{k'} k' p_{k'} \phi_{k'},\label{eq:mean1}
\end{align}
where we write $S_{k'}$ for the set of species with degree $k'$, and we have used that the degree distribution of the network can be written $p_k = N_k/N$ when $N\to \infty$, where $N_k$ is the number of species with degree $k$. We hence see that the mean of the first term in \cref{eq:cavity1} is non-vanishing in the thermodynamic limit.

Let us now examine the fluctuations of this same term [the first in \cref{eq:cavity1}]. One can see immediately from the approach in \cref{appendix:dmft} that the generating functional for the ensemble of all species factorises in the limit $N\to \infty$. This means that the variance of (or correlations between) any order parameters such as $\phi_{k}$ or $M_k$ are subleading in $1/N$ in the thermodynamic limit. So, keeping $A_{0j}$ and $\alpha_{0j}$ fixed, we see that fluctuations due to randomness in the wider community without the cavity species can always be neglected. Let us now examine the fluctuations of $\mu d^{-1}\sum_{j} A_{0j} \theta_j$ due to fluctuations in the interactions of the cavity species. Since the probability that each link in the network is independent of the rest of the links in the network, we have
\begin{align}
    \mathrm{Var}\left[\frac{ \mu}{d} \sum_{j} A_{0j}\theta_j \right] \approx \frac{\mu^2}{d^2}\sum_{j} \left[ \langle A_{0j} \rangle - \langle A_{0j} \rangle^2\right] \phi_{k_j}^2,
\end{align}
which is subleading in $1/d \sim 1/N$. So, we see that the first term in \cref{eq:cavity1} can be approximated by its mean in \cref{eq:mean1}. 

Let us now turn our attention to the third term in \cref{eq:cavity1}. We will see that in contrast to the first term, this term has non-vanishing fluctuations. We once again examine the ensemble average of this term (keeping the interaction coefficients with the cavity species fixed), noting again that the fluctuations of the order parameters of the wider community can be ignored. Using \cref{eq:deltaphi}, we find
\begin{align}
\left\langle \sum_{j}  A_{0j} \alpha_{0j} d\theta_j \right\rangle_0  &= \sum_{j}    A_{0j}\alpha_{0j} \langle d\theta_j \rangle_0 =  x_0 \sum_{j}    A_{0j} \alpha_{0j} \frac{d \phi_{k_j}}{dh}  \alpha_{j0}  ,\label{eq:mean3}
\end{align}
where we have used the fact that $A_{0j} = A_{j0}$.  

The expression obtained in \cref{eq:mean3} differs depending on the precise values of the interaction coefficients of the cavity species (noting that $x_0$ also depends on these quantities through \cref{eq:statvalue}). However, we can demonstrate that the sum over $j$ in \cref{eq:mean3} is a self-averaging quantity that can be replaced by its mean, meaning that all the relevant variation in the third term in \cref{eq:cavity1} can be captured by $x_0$, multiplied by a constant factor.

That is, we have
\begin{align}
\left\langle\sum_{j}      A_{0j} \alpha_{0j} \frac{d \phi_{k_j}}{dh}  \alpha_{j0} \right\rangle &= \frac{k \gamma \sigma^2}{d^2}\sum_{k'} k' p_{k'} \frac{d\phi_{k'}}{dh}, \nonumber \\
\mathrm{Var}\left[\sum_{j}      A_{0j} \alpha_{0j} \frac{d \phi_{k_j}}{dh}  \alpha_{j0} \right] &= \sum_{j}  \left[ \left\langle A_{0j} \right\rangle \left\langle (\alpha_{0j} \alpha_{j0} )^2\right\rangle -  \left\langle A_{0j} \right\rangle^2 \left\langle \alpha_{0j} \alpha_{j0} \right\rangle^2\right]  \left[\frac{d \phi_{k_j}}{dh}\right]^2.
\end{align}
We see once again that the fluctuations of this quantity vanish in the thermodynamic limit. We can thus approximate the sum $\sum_{j}      A_{0j} \alpha_{0j} \frac{d \phi_{k_j}}{dh}  \alpha_{j0}$ by its average. The third term in \cref{eq:cavity1} is thus well-approximated by 
\begin{align}
    \sum_{j}   A_{0j} \alpha_{0j} \delta\theta_j \approx x_0\frac{k \gamma \sigma^2}{d^2}\sum_{k'} k' p_{k'} \frac{d\phi_{k'}}{dh}, \label{eq:mean2}
\end{align}
where we see that the randomness is all accounted for by the variable $x_0$. In a certain sense, we were `lucky' that we could encapsulate the relevant fluctuations of the third term in \cref{eq:cavity1} entirely in the random variable $x_0$. The second term in \cref{eq:cavity1} is more complicated. To understand why, we compare with the cavity calculation that could have been performed to obtain the results in \cref{eq:stationary_abundance,eq:fixed_point_equations} (instead of the generating functional approach of \cref{appendix:dmft}).

\subsection{Lemma: relating the random variable \texorpdfstring{$z$ in \cref{eq:stationary_abundance}}{} to the interaction coefficients }
The fixed point satisfies of the dynamics in \cref{eq:glv_dynamcis} is given by 
\begin{align}
    x_i(1 - x_i + \sum_j A_{ij} \alpha_{ij} x_j + h_i) = 0.
\end{align}
Introducing a new species $0$ as a `cavity', one finds
\begin{align}
    x_j \approx x_j^{(0)} + \frac{d x_j}{dh_j} A_{j0} \alpha_{j0} x_0 .
\end{align}
One thus arrives at
\begin{align}
    x_0 \left(1 - x_0 +\frac{\mu}{d}\sum_{j} A_{0j} x_j^{(0)}+ \sum_{j} A_{0j} a_{0j} x_j^{(0)} + x_0\sum_{j} A_{0j} \alpha_{0j} \alpha_{j0} \frac{d x_j}{dh_j}\right)   = 0,
\end{align}
and consequently
\begin{align}
      x_0 = \max\qty(0, \frac{1 + \frac{\mu}{d}\sum_{j} A_{0j} x_j^{(0)}+ \sum_{j} A_{0j} a_{0j} x_j^{(0)} + h_j}{1 - \sum_{j} A_{0j} \alpha_{0j} \alpha_{j0} \frac{d x_j}{dh_j}}). \label{eq:fpnoaverage}
\end{align}
Supposing that species $0$ has original degree $k$, we can compare to \cref{eq:stationary_abundance}, and we see that the term $\sum_{j} A_{0j} a_{0j} x_j^{(0)}$ in the expression above corresponds to a Gaussian random variable, i.e.
\begin{align}
    \sum_{j} A_{0j} a_{0j} x_j^{(0)} = z\sigma \sqrt{\frac{k}{d}\sum_{k'}\frac{k' p_{k'}}{d}q_{k'}}.\label{eq:zcavity}
\end{align}
where $z$ is a zero-mean, unit-variance Gaussian random variable, as in \cref{eq:stationary_abundance}. One notes that we can can also deduce this from the cavity approach simply by computing the mean and the variance of $\sum_{j} A_{0j} a_{0j} x_j^{(0)}$. The variance is given as follows
\begin{align}
    V_x \equiv \left\langle \sum_{j} A_{0j} a_{0j} x_j^{(0)} \sum_{j'} A_{0j'} a_{0j'} x_{j'}^{(0)}\right\rangle  = \left\langle \sum_{j j'} \delta_{jj'} \frac{ \sigma^2}{d} A_{0j}  [x_j^{(0)} ]^2 \right\rangle  \approx  \frac{\sigma^2 k}{d^2} \sum_{k'}   k'  p_{k'} q_{k'} .
\end{align}
The second term in \cref{eq:cavity1} (i.e. $\sum_{j} A_{0j} a_{0j} \theta_j^{(0)}$) has a similar structure to the term $\sum_{j} A_{0j} a_{0j} x_j^{(0)}$. We see that it too must be a Gaussian random variable, with some correlation with the random variable $z$ that appears in \cref{eq:stationary_abundance} (given that it is also dependent on the same random variables $a_{0j}$). If we can find the variance of $\sum_{j} A_{0j} a_{0j} \theta_j^{(0)}$ and its correlation with $z$, then we will understand fully how to relate $\mu_0^{(k)}$ to the fixed point quantities in \cref{eq:fixed_point_equations}, given that we already have the approximations for the first and third terms in \cref{eq:cavity1} in \cref{eq:mean1,eq:mean2} respectively.

\subsection{Relating the fluctuating part of \texorpdfstring{\cref{eq:cavity1} to the fluctuating part of \cref{eq:stationary_abundance}}{}}

Let us now compute the variance of the quantity $\sum_{j} A_{0j} a_{0j} \theta^{0}_j$ in \cref{eq:cavity1}. We find
\begin{align}
    V_\theta \equiv \left\langle \sum_{j} A_{0j} a_{0j} \theta_j^{(0)} \sum_{j'} A_{0j'} a_{0j'} \theta_{j'}^{(0)}\right\rangle  = \left\langle \sum_{j j'} \delta_{jj'} \frac{ \sigma^2}{d} A_{0j}  [\theta_j^{(0)} ]^2 \right\rangle  =  \frac{ \sigma^2 k}{d^2} \sum_{k'}   k'  p_{k'} \phi_{k'}  .
\end{align}
We can thus think of the second term in \cref{eq:cavity1} as also being a zero-mean Gaussian random variable, so we write
\begin{align}
    \sum_{j} A_{0j} a_{0j} \theta^{0}_j \equiv y\sqrt{V_\theta},\label{eq:ydef}
\end{align}
where $y$ is a Gaussian random variable with unit variance. Let us now understand how $y$ is related to $z$ in \cref{eq:zcavity} by finding the covariance
\begin{align}
    C_{x\theta} \equiv \left\langle \sum_{j} A_{0j} a_{0j} \theta_j^{(0)} \sum_{j'} A_{0j'} a_{0j'} x_{j'}^{(0)}\right\rangle  = \left\langle \sum_{j j'} \delta_{jj'} \frac{ \sigma^2}{d} A_{0j}  \theta_j^{(0)} x_j^{0} \right\rangle  =  \frac{ \sigma^2 k}{d^2} \sum_{k'}   k'  p_{k'} M_{k'}  .
\end{align}
We can thus write
\begin{align}
    y = \frac{C_{x\theta}}{\sqrt{V_x V_\theta}} z + z'\sqrt{1 - \frac{C_{x\theta}}{\sqrt{V_x V_\theta}}} , \label{eq:yrelationtoz}
\end{align}
where $z'$ is a zero-mean unit-variance Gaussian random variable that is independent of $z$. We are now in a position to write \cref{eq:cavity1} entirely in terms of the statistics of the surviving community.

\subsection{Incoming and outgoing statistics of nodes with given degree}

Now, inserting \cref{eq:mean1,eq:mean2,eq:ydef,eq:yrelationtoz} into \cref{eq:cavity1}, we obtain
\begin{align}
    \mu_0^{(k)} = \frac{\mu k}{d^2}\sum_{k'} k' p_{k'} \phi_{k'} + z_k \frac{\sigma \sqrt{k}}{d}  \frac{\sum_{k'} k' p_{k'} M_{k'}}{\sqrt{\sum_{k'} k' p_{k'} q_{k'}}} + C_{z'} z' +x_k(z_k) \frac{\gamma \sigma^2 k}{d^2}\sum_{k'} k' p_{k'} \frac{d \phi_{k'}}{dh}. \label{eq:mu0k}
\end{align}
where we have now evaluated some terms explicitly to highlight their dependence on $k$, and we simply write $C_{z'}$ for the coefficient multiplying $z'$, since this will not affect the quantities in which we are interested.

Let us now consider the following quantity 
\begin{align}
\nu_0^{(k)} = \sum_{j} A_{j0} \alpha_{j0} \theta_j ,
\end{align}
which instead tells us about the outgoing links of a node with degree $k$. We can perform exactly the same manipulations as we did for $\mu_0^{(k)}$ to arrive at
\begin{align}
    \nu_0^{(k)} = \frac{\mu k}{d^2}\sum_{k'} k' p_{k'} \phi_{k'} + z_k \frac{\gamma \sigma \sqrt{k}}{d}  \frac{\sum_{k'} k' p_{k'} M_{k'}}{\sqrt{\sum_{k'} k' p_{k'} q_{k'}}} + C_{z'}'z' +x_k(z_k) \frac{ \sigma^2 k}{d^2}\sum_{k'} k' p_{k'} \frac{d\phi_{k'}}{dh}. \label{eq:nu0k}
\end{align}
We notice the symmetry between the expressions in \cref{eq:mu0k,eq:nu0k}. What was an effect of the neighbours of a node on the node itself in \cref{eq:mu0k} becomes the effect of the node on its neighbours in \cref{eq:nu0k}. This is why we see factors of $\gamma$ multiplying complementary terms in the two expressions. 

Now, to obtain the ensemble average of the above expressions, we simply average over realisations of the variable $z_k$, conditioning on the survival of the cavity species $0$. This means that require $x_k(z_k)>0$, which in turn requires that
\begin{align}
    z_k > -\Delta_k(h) \equiv  -\frac{1 + \mu k\sum_{k'}p_{k'}k'M_{k'}/d^2 + h}{\sigma\sqrt{k\sum_{k'}p_{k'}k'q_{k'}}/d} .
\end{align}
The probability of survival is given by
\begin{align}
    \phi_k = \int_{-\Delta_k}^\infty \dd{z} \frac{1}{\sqrt{2\pi}} e^{-z^2/2}.
\end{align}
Hence, averaging over the variable $z_k$ in \cref{eq:mu0k,eq:nu0k}, we obtain
\begin{align}
    \frac{\mu_{k}^{\star\mathrm{in}}}{d^\star} \equiv \frac{\langle\mu_0^{(k)} \theta_0\rangle }{ kr } &= \frac{\mu}{d} +  \frac{\sigma^2}{d}  \dv{\phi_{k}}{h}  \frac{\sum_{k'} k' p_{k'} M_{k'}}{\sum_{k'} k' p_{k'} \phi_{k'}}  + \frac{\gamma \sigma^2}{d}  M_k \frac{\sum_{k'} k' p_{k'} \dv{\phi_{k'}}{h}}{\sum_{k'} k' p_{k'} \phi_{k'}} , \nonumber \\
    \frac{\mu_{k}^{\star\mathrm{out}}}{d^\star}\equiv \frac{\langle\nu_0^{(k)} \theta_0\rangle }{kr } &= \frac{\mu}{d} + \frac{\gamma\sigma^2}{d}  \dv{\phi_{k}}{h} \frac{\sum_{k'} k' p_{k'} M_{k'}}{\sum_{k'} k' p_{k'} \phi_{k'}} + \frac{ \sigma^2 }{d}M_k \frac{\sum_{k'} k' p_{k'} \dv{\phi_{k'}}{h}}{\sum_{k'} k' p_{k'} \phi_{k'}}, \label[plural_equation]{appendix:eq:mu_inout_k}
\end{align}
where here we have used the fact that 
\begin{align}
    \int_{-\Delta}^\infty \dd{z} e^{-z^2/2} z = \frac{1}{\sqrt{2\pi}} e^{-\Delta^2/2} = \dv{\phi_{k}}{h} \sqrt{\sigma^2 \frac{k}{d^2} \sum_{k'}p_{k'} k' q_{k'}},
\end{align}
and we recall the definitions of $r$ from the main text
\begin{align}
    r = \frac{\sum_kp_k \phi_k k}{\sum_kp_k}.
\end{align}
To express the incoming and outgoing interaction strengths in \cref{appendix:eq:mu_inout_k} in terms of the degree in the surviving community $k^*$ (rather than the degree in the initial community $k$), we use the correspondence $\E[k^* \mid k] = rk$ [see \cref{section:surviving_network} for a discussion and \cref{appendix:surviving_network} for mathematical details], as well as the fact that the expected degrees in the surviving community are concentrated around their mean value. Hence, if we treat $k$ as a continuous variable, we can approximate $k^* = rk$ to leading order in $1/d$.

Practically, the curves in \cref{fig:interaction_degree_correlations} are produced by plotting $\mu_{k}^{\star\mathrm{in}}/d^\star$ in \cref{appendix:eq:mu_inout_k} against $k/r$. This is equivalent to plotting $\mu_{k^\star}^{\star\mathrm{in}}/d^\star$ against $k^\star$. 

\subsection{Interpretation of the trends in \texorpdfstring{\cref{fig:interaction_degree_correlations}}{}}

Let us now consider how the cavity approach that we have taken can help us to understand the trends in \cref{fig:interaction_degree_correlations}. This is accomplished by interpreting physically each of the terms in \cref{eq:mu0k,eq:nu0k}, with the help of \cref{eq:cavity1}. 

\cref{eq:mu0k} describes the sum of incoming interactions to a node of original degree $k$ as a random variable. The first term in this expression is deterministic, and is simply the mean interaction, weighted by the number of surviving neighbours (which will depend on $k$). The second and third terms encode the fluctuations in the weights of the neighbours' interactions. However, we note that once we average over the disorder (conditioning on survival of species $0$), the term proportional to $z'$ vanishes, and the Gaussian distribution of $z_k$ is truncated. That is, only species with sufficiently favourable interactions survive, and this biases the mean interaction of surviving species towards higher values. Finally, the last term encapsulates the fact the survival of the neighbours of a species is dependent on the abundance of that species. In turn, the survival of the neighbouring species affects the statistics of the incoming interactions (i.e. the abundance of a species affects its own incoming interactions via its effect on its neighbours). We note that this last effect depends on the correlation between the incoming and outgoing links.

In the case where $\gamma = 0$ (i.e. there is no correlation between the incoming interactions to a species and outgoing effect of a species on its neighbours, as is the case in \cref{fig:interaction_degree_correlations}b), the final term mentioned above does not contribute. Instead, only the direct influence of a species' neighbours is relevant. Since we condition on the survival of the species with degree $k$, the incoming interactions cannot be too negative. This is encapsulated by the lower limit imposed on the truncated Gaussian random variable $z_k$ in \cref{eq:mu0k}. For small $k$, this lower limit is effectively $-\infty$, which is reflected in the survival of nearly all species with small degree (i.e. $\phi_k \approx 1$ for small $k$ see \cref{fig:phik_and_Mk_correlations}). This means that the term involving $z_k$ averages to nil when we integrate over all its possible values, and we find that species with low degree have interactions that are the same as the original community. However, as we increase $k$, $\phi_k$ decreases, and the lower limit on the integration of $z_k$ increases also. This means that a bias is introduced, whereby only species with more favourable incoming interactions survive. This explains the upwards trend in \cref{fig:interaction_degree_correlations}b for the incoming interactions. 

Likewise, we can interpret each of the terms in \cref{eq:nu0k} as follows: The first is again simply the mean interaction, weighted by the number of surviving neighbours. The second and third terms now reflect that each outgoing interaction from a node can fluctuate, but these outgoing interactions correlate with the incoming interactions. For this reason, the outgoing interactions can once again be related to the variable $z_k$, and thus the survival probability of the species with original degree $k$. Since we condition on this survival, this biases the outgoing interactions so that the incoming interactions are favourable (note that resulting effect on the outgoing interactions then depends on the sign of $\gamma$). Finally, the last term again encapsulates the fact that whether or not the neighbours of a species survive is dependent on the abundance of that species. Since we only look at the outgoing interactions from a species (with degree $k$) to species that survive, if the abundance of the species with degree $k$ is higher, its influence on the survival of the surrounding species is greater. For greater abundances, a greater number of species can be killed, and the correction to the average outgoing interaction is greater. Since, $M_k$ reduces with increasing $k$, we see that the effect of this term is greatest for small $k$, and it reduces to nil for large $k$. This explains the downwards trend in the outgoing interactions in \cref{fig:interaction_degree_correlations}b.

The case of $\gamma = 0$ in \cref{fig:interaction_degree_correlations}b is useful, because it separates the dependence of the outgoing and incoming interactions. We see that by varying $\gamma$, we simply obtain a superposition of the aforementioned effects. For example, when $\gamma = 1$, the outgoing and incoming interactions must be the same. Hence, $\mu_k^\mathrm{out} = \mu_{k}^\mathrm{in}$ and the corresponding curve in \cref{fig:interaction_degree_correlations}a is seen to be a kind of `average' of the upwards and the downwards trends. On the other hand, for $\gamma = -1$, we see that the fluctuations in the in- and out-interactions are exactly the negative of each other, hence the kind of mirror-image effect in \cref{fig:interaction_degree_correlations}c.

\section{Structure of the surviving network}\label{appendix:surviving_network}
\subsection{Probability that species with degree \texorpdfstring{$k$ and $k'$}{} interact in the surviving community}\label{appendix:prob_of_ij_interacting_surviving}
To find the statistics of the adjacency matrix in the surviving community, we follow a strategy employed in \cite{baronBreakdownRandomMatrixUniversality2023b} that was used to find the statistics of the surviving interaction matrix $\vb*\alpha$ in the fully connected model. Consider the following modification of the generating functional in \cref{appendix:dmft}, which includes an additional term proportional to the interaction matrix in the surviving community 
\begin{align}
      Z[\vb*\lambda] = \int\DD{\vb x}\DD{\wh{\vb x}}Z_0[\vb x, \wh{\vb x}, \vb* \psi=0]\exp[-i\sum_{ij}\int\dd{t}A_{ij}\alpha_{ij}\wh x_i(t) x_j(t) + i\sum_{ij}A_{ij}\int\dd{t}\lambda_{ij}(t)\theta_i(t)\theta_j(t)].\label{appendix:eq:generating_functional_interaction}
\end{align}
The functional $Z_0[\vb x, \wh{\vb x}, \vb* \psi=0]$ is the remaining part of the generating functional which appears in \cref{appendix:eq:generating_functional}, it is not relevant to our arguments in this section. As in \cref{appendix:interaction_statistics}, $\theta_i(t)=1$ if the corresponding abundance $x_i(t) > 0$, and is zero otherwise. In other words, $\theta_i(t)$ is equal to $1$ only if species $i$ is alive at time $t$. The functions $\lambda_{ij}(t)$ are auxiliary fields which we will set to zero at the end of this derivation.

Taking a functional derivative of $Z$ with respect to $\lambda_{ij}(t)$, then setting $\lambda_{ij}(t) = 0$, yields
\begin{align}
      \fdv{Z[\vb*\lambda]}{\lambda_{ij}(t)} = iA_{ij}\theta_i(t)\theta_j(t),\label{appendix:eq:A_from_Z}
\end{align}
We are interested in the average interactions between species with degree $k$ and $k'$. Using \cref{appendix:eq:A_from_Z}, we can relate this quantity to the generating functional via
\begin{align}
      \frac{1}{N_k^*N_{k'}^*}\sum_{i\in S_k^*}\sum_{j\in S_{k'}^*}\avg{A_{ij}\theta_i^*\theta_j^*}_{\vb*\alpha, \vb A} = -i \lim_{t\to\infty}\frac{1}{N_k^*N_{k'}^*}\sum_{i\in S_k^*}\sum_{j\in S_{k'}^*}\eval{\avg*{\fdv{Z[\vb*\lambda]}{\lambda_{ij}(t)}}_{\vb*\alpha, \vb A}}_{\lambda=0}.
\end{align}
Averaging $Z[\vb*\lambda]$ over $\vb A$ and $\vb*\alpha$ proceeds similarly to the average calculated in \cref{appendix:dmft}, we find 
\begin{align}
      &\avg{Z[\vb*\lambda]}_{\vb A, \vb*\alpha} = \int\DD{\vb x}\DD{\wh{\vb x}}Z_0[\vb x, \wh{\vb x}, \vb* \psi=0]\nonumber \\
      &\times\exp[\frac{1}{2}\sum_{ij}\ln\qty{1 + \frac{k_ik_j}{dN}\qty(\avg*{e^{-i\int\dd{t}\qty\Big(\alpha\wh x_i(t) x_j(t) + \beta\wh x_j(t) x_i(t))}}_{(\alpha, \beta)}e^{i\int\dd{t}\qty\big(\lambda_{ij}(t) + \lambda_{ji}(t))\theta_i(t)\theta_j(t)} - 1)}], \label{appendix:eq:generating_functional_network_averaged}
\end{align}
where, similarly to in \cref{appendix:dmft}, we have written $\alpha_{ij}\to\alpha$ and $\alpha_{ji}\to\beta$ because the joint distribution of $(\alpha_{ij}, \alpha_{ji})$ does not depend on the indices $i,j$. Differentiating \cref{appendix:eq:generating_functional_network_averaged} with respect to $\lambda_{ij}(t)$ and setting $\lambda_{ij}(t) = 0$ gives 
\begin{align}
      \avg{A_{ij}\theta_i(t)\theta_j(t)}_{\vb A, \vb*\alpha} = \Bigg\langle~\frac{\frac{k_ik_j\theta_i(t)\theta_j(t)}{dN}\avg*{e^{-i\int\dd{t}\qty\Big(\alpha\wh x_i(t) x_j(t) + \beta\wh x_j(t) x_i(t))}}_{(\alpha, \beta)}}{1 + \frac{k_ik_j}{dN}\qty(\avg*{e^{-i\int\dd{t}\qty\Big(\alpha\wh x_i(t) x_j(t) + \beta\wh x_j(t) x_i(t))}}_{(\alpha, \beta)} - 1)}~\Bigg\rangle_{\vb A, \vb*\alpha}.\label{appendix:eq:avgA_lambda_not_zero}
\end{align}
This expression simplifies greatly if the network is dense (where the average degree $d$ is large), as it is in our model. From the statistics of the interactions $\alpha_{ij}$ in \cref{eq:interaction_statistics}, we know that $\alpha_{ij} = \order{d^{-1/2}}$ and $\avg{\alpha_{ij}}_{\vb*\alpha} = \order{d^{-1}}$. Hence, to leading order in $1/d$, the statistics of the interactions do not directly contribute to \cref{appendix:eq:avgA_lambda_not_zero} and we have
\begin{align}
      \avg{A_{ij}\theta_i(t)\theta_j(t)}_{\vb A, \vb*\alpha} = \avg*{\frac{k_ik_j}{dN}\theta_i(t)\theta_j(t)}_{\vb A, \vb*\alpha} + \order{d^{-1}}.
\end{align}
Averaging over species with common degree in the original community now gives 
\begin{align}
      \frac{1}{N^\star_kN^\star_{k'}}\sum_{i\in S_k^\star}\sum_{j\in S_{k'}^\star}\avg{A_{ij}\theta^\star_i\theta^\star_j}_{\vb A, \vb*\alpha} &= \frac{1}{N^\star_kN^\star_{k'}}\sum_{i\in S_k^\star}\sum_{j\in S_{k'}^\star}\avg*{\frac{k_ik_j}{dN}\theta_i(t)\theta_j(t)}_{\vb A, \vb*\alpha}, \nonumber \\
      &= \frac{kk'}{dN}\label{appendix:eq:surviving_Aij}
\end{align}
as claimed in the main text. 

The same method can be used to compute any statistics of the adjacency matrix in the surviving community. As we need it in the following section, we also have 
\begin{align}
    \frac{1}{N_k^\star N_{k'}^\star N_{k''}^\star}\sum_{i\in S^*_k}\sum_{j \in S^*_{k'}}\sum_{l\in S^*_{k''}}\avg{A_{ij}A_{il}\theta_i^\star \theta_j^\star\theta_l^\star}_{\vb A, \vb*\alpha} = \frac{k^2k'k''}{d^2N^2}.\label{appendix:eq:AA_stats_surviving_community}
\end{align}
All higher moments of the adjacency matrix have similarly simple forms.
\subsection{Degree sequence in the surviving community}\label{appendix:degree_sequence}
Here we detail the derivation of the degree sequence in the surviving community. First, we will show that the expected degree of a species in the surviving community, given its degree in the original community, is given by \cref{eq:surviving_degree_sequence} in the main text. We will then show that the degrees concentrate around their mean value. 

To compute $\E[k^*_i\mid k_i, i\in S^*]$, we write it in terms of the adjacency matrix in the surviving community. This gives [using \cref{appendix:eq:surviving_Aij}]
\begin{align}
    \E[k^*_i \mid k_i, i\in S^*] &= \E\qty[\sum_{j\in S^*} A^*_{ij} \mid k_i, i\in S^*], \nonumber \\
    &= \sum_{k'}N^*_{k'}\qty(\frac{k_ik'}{dN} + \order{d^0}), \nonumber \\
    &= k_i r + \order{d^0},\label{appendix:eq:expected_surviving_degree}
\end{align}
where we recall that $r = \sum_k p_k k\phi_k/\sum_k p_k k$ is the average neighbour survival rate in the community. The calculation of the variance proceeds in the same way, it relies on the additional calculation in \cref{appendix:eq:AA_stats_surviving_community}. We have 
\begin{align}
    \E[(k^*_i)^2 \mid k_i, i\in S^*] - \E[k^*_i \mid k_i, i\in S^*]^2 &= \E\qty[\sum_{jl\in S^*} A^*_{ij}A^*_{il} \mid k_i, i\in S^*] - \E\qty[\sum_{j\in S^*} A^*_{ij} \mid k_i, i\in S^*]^2, \nonumber \\
    &= \sum_{kk'}N^*_kN^*_{k'}\frac{(k_i)^2kk'}{d^2N^2} - (k_ir)^2 + \order{d^0}, \nonumber \\
    &= \order{d^0},
\end{align}
as claimed in the main text. 
\subsection{Degree distribution in the surviving community}\label{appendix:degree_distribution}
To leading order in $1/d$, the degrees of species in the surviving community concentrate around the mean value in \cref{appendix:eq:expected_surviving_degree}. Using this, we can find an expression for the degree distribution which is accurate to leading order in $1/d$ by simply approximating the degree sequence of a species in the surviving community with $k^\star = kr$, where $k$ is the degree of the species in the original community. As $kr$ is not in general an integer, we will find an expression for the function $P^\star(k^\star/M)= Mp^\star_{k^\star}$, where $M$ is the number of different degrees in the community (the number of distinct values of $k_i$). We also note that when $N$ is large, the degree distribution in the original community can be written similarly as $P(k/M) = Mp_k$. These expressions are normalised and satisfy 
\begin{align}
    \frac{1}{N}\sum_if(k_i) &= \sum_k p_k f(k) \approx \int_{\kappa_\text{min}}^{\kappa_\text{max}}P(\kappa) F(\kappa) \dd{\kappa}, \nonumber \\
    \frac{1}{N^\star}\sum_{i\in S^\star} f(k_i^\star) &= \sum_{k^\star}p^\star_{k^\star}f(k^\star) \approx\int_{\kappa_\text{min}^\star}^{\kappa_\text{max}^\star}P^\star(\kappa) F(\kappa) \dd{\kappa},\label{appendix:eq:continuous_degree_distributions}
\end{align}
where $f$ is an arbitrary function and $F$ satisfies
\begin{align}
    \frac{1}{M}F\qty(\frac{k}{M}) &= f(k), \nonumber\\
\end{align}
and where $\kappa = K/M$ and $\kappa^\star = k^\star/M = r\kappa$. The approximations in \cref{appendix:eq:continuous_degree_distributions} hold for large $M$, which allows us to approximate the sums as integrals.

To find an expression for the degree distribution in the surviving community, we observe that the second of the expressions in Eqs.~(\ref{appendix:eq:continuous_degree_distributions}) can also be computed as follows (using the approximation $k^\star = kr$)
\begin{align}
      \frac{1}{N^\star}\sum_{i\in S^\star}f(k^\star_i) &= \frac{1}{N^\star}\sum_k N^\star_k f(kr) \nonumber \\
      &= \frac{1}{\phi}\sum_k p_k \phi_k f(kr)\nonumber \\
      &= \frac{1}{\phi}\int_{\kappa_{\text{min}}}^{\kappa_\text{max}}\dd{\kappa} p(\kappa) \phi(\kappa)F(\kappa r) \nonumber \\
      &= \frac{1}{\phi r}\int_{\kappa^\star_{\text{min}}}^{\kappa^\star_\text{max}}\dd{\kappa} p\qty(\frac{\kappa}{r}) \phi\qty(\frac{\kappa}{r})F(\kappa)
\end{align}
where, similarly to the functions $f$ and $F$, we have written $\phi(k/M) = M\phi_k$. The two expressions $\frac{1}{N^\star}\sum_{i\in S^\star} f(k_i^\star)$ must be must be equal. and as the function $f$ is arbitrary, we conclude that the following functions are be equal 
\begin{align}
      p^\star(\kappa) = \frac{1}{\phi r}p\qty(\frac{\kappa}{r}) \phi\qty(\frac{\kappa}{r}),
\end{align}
which is \cref{eq:surviving_degree_distribution} in the main text.

\end{document}